\documentclass[aps,prd,twocolumn,nofootinbib,groupedaddress,amsfonts,floatfix]{revtex4}

\usepackage{graphicx}
\usepackage{setspace}
\usepackage{pifont}
\usepackage{amsmath}
\usepackage{dcolumn}

\begin{document}

\title{Systematic and statistical errors in a bayesian approach to the estimation of the neutron-star equation of state using advanced gravitational wave detectors}

\author{Leslie Wade${}^1$} 
\author{Jolien D. E. Creighton${}^1$}
\author{Evan Ochsner${}^1$}
\author{Benjamin D. Lackey${}^2$}
\author{Benjamin F. Farr${}^3$}
\author{Tyson B. Littenberg${}^3$}
\author{Vivien Raymond${}^4$}

\affiliation{${}^1$Department of Physics, University of Wisconsin - Milwaukee, P.O. Box 413, Milwaukee, Wisconsin 53201} 
\affiliation{${}^2$Department of Physics, Princeton University, Princeton, NJ 08544, USA} 
\affiliation{${}^3$Center for Interdisciplinary Exploration and Research in Astrophysics (CIERA) \& Dept. of Physics and Astronomy, 2145 Sheridan Rd, Evanston, IL 60208}
\affiliation{${}^4$LIGO - California Institute of Technology, Pasadena, CA 91125, USA}

\begin{abstract}
Advanced ground-based gravitational-wave detectors are capable of measuring tidal influences in binary neutron-star systems.  In this work, we report on the statistical uncertainties in measuring tidal deformability with a full Bayesian parameter estimation implementation.  We show how simultaneous measurements of chirp mass and tidal deformability can be used to constrain the neutron-star equation of state.  We also study the effects of waveform modeling bias and individual instances of detector noise on these measurements.  We notably find that systematic error between post-Newtonian waveform families can significantly bias the estimation of tidal parameters, thus motivating the continued development of waveform models that are more reliable at high frequencies.
\end{abstract}


\maketitle
 

 
\section{Background and Motivation}

Advanced interferometric gravitational-wave (GW) detectors currently under construction are expected to begin operating in the next few years.  Advanced LIGO \cite{aLIGO} is expected to achieve its design sensitivity c.~2019 \cite{ObservingScenarios}, at which time the detection rate of binary neutron-star (BNS) events in a single detector is expected to be $\sim$40 yr$^{-1}$, though this value is quite uncertain and ranges from 0.4--400 yr$^{-1}$ \cite{LIGO_Rates}.

When a compact binary coalescence (CBC) signal is detected \cite{LowMass_Search_BigDog,HighMass_Search}, the corresponding interferometer data stream segment is sent through a parameter estimation pipeline to determine the source parameters of the system.  Some of these source parameters include the binary component masses and spins, the sky location, distance, and orientation of the system.  Bayesian inference is used to explore the probability distribution of the CBC's source parameters by comparing model waveform templates, whose form depends on these source parameters, to the data stream segment containing the GW.  For this work, we use \texttt{lalinference\char`_mcmc}, which is included in the \texttt{LALInference} LSC Algorithm Library \cite{LAL}, as our parameter estimation pipeline.  It is a Markov Chain Monte Carlo (MCMC) sampler designed to efficiently explore the full waveform parameter space in order to make reliable and meaningful statements about CBC source parameters \cite{MCMC1,MCMC2,S6PE}.

This paper's focus is on measuring the effect of tidal influence on BNS GW signals with advanced detectors.  Neutron stars (NSs) in merging CBC systems will be tidally deformed by the gravitational gradient of their companion across their finite diameter.  This effect is insignificant at large separations but becomes increasingly significant as the NSs near each other \cite{Flanagan_Hinderer}.  The internal structure of a NS, which is characterized by its equation of state (EOS), determines how much each star will deform.  The amount that a NS deforms will affect the orbital decay rate, which is encoded in the observed gravitational waveform.  Therefore, if a gravitational signal from a BNS system is detected, then such a detection could provide insight into the NS EOS \cite{Flanagan_Hinderer,Hinderer_Lackey,Read_NR,Pannarale_EOS}.

In order to make meaningful statements regarding the recoverability of tidal parameters from BNS signals, it is important to understand the effects of error on parameter estimation.  One such obstacle to measuring tidal influence is accurate waveform modeling.  The error resulting from inaccurate waveform models is a kind of systematic error.  Some of the most commonly used CBC waveforms rely on a post-Newtonian (PN) expansion in orbital speed.  As the CBC inspirals, the orbital speed of the binary components increases leading to a higher frequency signal.  These waveform families are thus unreliable at high frequencies where orbital speeds become large \cite{BIOPS} and tidal effects emerge.  Another difficulty in measuring tidal influence results from fluctuations in detector noise.  This type of error is called statistical error.  Tidal influences only noticeably affect the final high frequency orbits of the binary where the detector noise (in strain units) is comparatively large.  Extracting such a small influence occurring in the high frequency band is an investigation at the very brink of our detectors' sensitivity.  Even small fluctuations in detector noise might be able to dramatically affect the recovery of tidal deformability.  Understanding the magnitude of these two sources of error is the core motivator of this work.

Several studies have used the Fisher Information Matrix (FM), which is only valid in the large signal-to-noise ratio (SNR) limit, to estimate the measurability of tidal effects on the CBC gravitational waveform \cite{Flanagan_Hinderer,Hinderer_Lackey,Read_NR,Read_ME,Lackey_NSBH,Damour_EOB,Maselli}.  Flanagan and Hinderer \cite{Flanagan_Hinderer} were among the first to show that advanced detectors can constrain the tidal influence of NSs on the early inspiral portion of the CBC waveform.  They notably use PN waveforms truncated at 400 Hz to remove the unreliable high-frequency portion of the PN model.  Hinderer {\it et al.}\ \cite{Hinderer_Lackey} later investigated how well constraints on the tidal deformability from the early inspiral can discriminate between several theoretical NS EOSs.  Also using PN waveforms, they find that advanced detectors will likely only be able to probe stiff EOSs.

Further FM studies moved away from the use of PN waveforms in favor of waveforms that are more reliable at high frequencies.  Read {\it et al.}\ \cite{Read_NR,Read_ME} probed the late inspiral portion of the BNS waveform with numerical relativity (NR) simulations, which are accurate during the late inspiral and merger epochs.  They find that the additional high frequency information results in greater measurement accuracy of the tidal deformability.  Damour, Nagar, and Villain \cite{Damour_EOB} also probed beyond the early inspiral with tidally corrected effective-one-body (EOB) waveforms, which they claim to be accurate up to merger.  They show that advanced detectors should in fact be able to constrain the NS EOS for reasonably loud signals.

While the above mentioned studies are informative, the FM is not always trustworthy in estimating the measurability of source parameters \cite{Vallisneri, Carl_FM, EFM1, EFM2}.  Though it is known that FM estimates are only accurate for loud signals, recent investigations have highlighted additional shortcomings of FM estimates when compared to real GW parameter estimation pipelines \cite{Carl_FM}.  It is now clear that there is no substitute for full Bayesian results when making definitive statements regarding parameter estimation.

Del Pozzo {\it et al.}\ \cite{DelPozzo} recently performed Bayesian simulations of BNS systems with a tidally corrected PN waveform.  They find that advanced detectors will be able to measure tidal effects on GW signals and constrain the NS EOS by combining information from many BNS sources.  While this result is very important, their analysis assumes that true BNS signals have the exact same form as their model.  Although the authors acknowledge this limitation, it is necessary to study how much their result depends on this assumption.

Recently, there have been several FM investigations that have studied the effects of systematic error on the measurability of tidal parameters \cite{Yagi_Yunes,Favata,Read_ME,Lackey_NSBH}.  In particular, Yagi and Yunes in \cite{Yagi_Yunes} and Favata in \cite{Favata} both find that current PN waveforms, which are known only up to 3.5PN order \cite{BIOPS}, cannot be used to make accurate measurements of tidal effects.  This is an extremely important result that motivates a full Bayesian investigation into the effect of systematic error from tidally corrected PN waveforms on parameter estimation.

In this work, we use a full Bayesian framework to demonstrate the ability of advanced detectors to constrain the NS EOS by measuring the effects of tidal influence on BNS signals.  We estimate the anticipated measurement uncertainty associated with using the advanced LIGO/Virgo network \cite{aLIGO,Virgo} to recover tidal influence in BNS systems.  We find that systematic error inherent in the current PN inspiral waveform families significantly biases the recovery of tidal parameters.  Additionally, we find that individual instances of detector noise can on occasion considerably reduce the measurability of tidal parameters.  We consider only BNS systems.

This work is organized as follows.  In Sec.~\ref{Sec_CBC_PN} we review how tidal influences affect the CBC waveform.  In Sec.~\ref{Sec_PE} we briefly outline the parameter estimation pipeline used in this analysis and present measurement uncertainty estimates for the recovery of tidal influences in BNS systems.  In Sec.~\ref{Sec_NS_EOS} we explain how simultaneous mass-like and radius-like measurements, specifically the measurement of chirp mass and tidal deformability, can help constrain the NS EOS.  In Sec.~\ref{Sec_Error} we describe the two main sources of error in parameter estimation and how much each source of error affects the recovery of tidal parameters.  We finish with a summary of our main results in Sec.~\ref{Sec_Conclusion}.  We also refer the interested reader to Appendix~\ref{PN_derivations} where we derive how the tidal corrections appear in several PN waveform families.


\section{Tidal Corrections to CBC PN Waveform Families}
\label{Sec_CBC_PN}

In this section, we review the effects of tidal influences on the CBC waveform.  For a more complete discussion, refer to Appendix~\ref{PN_derivations}, which outlines how tidal effects appear in the following PN waveform families: TaylorT1, TaylorT2, TaylorT3, TaylorT4, and TaylorF2.  For more details regarding each of these waveform families, see \cite{BIOPS} and references therein.

\subsection{Constructing tidally corrected PN waveforms}

To model the CBC waveform, it is customary to approximate each massive body as having infinitesimal size.  As the two bodies orbit, GWs carry energy away from the system causing their separation to decrease and their orbital frequency to increase.  The energy and luminosity of this point-particle system ($E_{\rm pp}$ and ${L}_{\rm pp}$ respectively) are currently known to 3.5 post-Newtonian (PN) order\footnote{The energy has recently been calculated to 4PN order \cite{4PN_Energy}.} \cite{BIOPS}. 

If the two compact objects are NSs, each will start to deform under the tidal field of the other as their separation decreases.  The deformation of each body will have an effect on the rate at which the bodies coalesce.  BNS systems therefore depart from the point-particle approximation at high frequencies and require an additional correction to the energy and luminosity of the system relative to the point-particle terms.

Since a NS in a binary system will deform under the tidal influence of its companion, its quadrupole moment ${\cal Q}_{ij}$ must be related to the tidal field ${\cal E}_{ij}$ caused by its companion.  For a single NS, to leading order in the quasi-stationary approximation and ignoring resonance,
\begin{equation}
{\cal Q}_{ij}=-\lambda{\cal E}_{ij},
\end{equation}
where $\lambda=(2/3)k_2 R^5/G$ parameterizes the amount that a NS deforms \cite{Flanagan_Hinderer}.  The $i$ and $j$ are spatial tensor indices, $k_2$ is the second Love number, and $R$ is the NS's radius.  Since $\lambda$ parameterizes the severity of a NS's deformation under a given tidal field, it must depend on the NS EOS.  NSs with large radii will more easily be deformed by the external tidal field, because there will be a more extreme gravitational gradient over their radius.  For a fixed mass, NSs with large radii are also referred to as having a stiff EOS, and, for the same mass, NSs with small radii have a soft EOS.  Therefore, NSs that have large values of $\lambda$ will have large radii, a stiff EOS, and become severely deformed in BNS systems; on the other hand, NSs that have small values of $\lambda$ will have small radii, a soft EOS, and will be less severely deformed in these systems.

Tidal effects are most important at small separations and therefore at high frequencies in BNS systems.  Tidal corrections to the energy $\delta E_{\rm tidal}$ and tidal corrections to the luminosity $\delta{L}_{\rm tidal}$ add linearly to the point-particle energy $E_{\rm pp}$ and luminosity ${L}_{\rm pp}$.  Though the leading order tidal correction is a Newtonian effect, it is often referred to as a 5PN correction, because it appears at 5PN order relative to the leading order point-particle term.  In this work, we keep the leading order (5PN) and next-to-leading order (6PN) corrections to the energy and luminosity \cite{VinesFlanaganHinderer}:
\begin{widetext}
\begin{eqnarray}
\label{CBC_dE}
\delta{E}_{\rm tidal}&=&-\frac{1}{2}c^2M\eta x\left[   -\left(\frac{9}{\chi _1}-9\right)\frac{c^{10}}{G^4}\frac{\lambda_1}{M^5}x^{5} - \left(\frac{33}{2 \chi _1}-\frac{11}{2}+\frac{11}{2}\chi _1-\frac{33}{2}\chi _1^2\right)\frac{c^{10}}{G^4}\frac{\lambda_1}{M^5}x^{6} + (1\longleftrightarrow 2)   \right]\\
\label{CBC_dEdot}
\delta{L}_{\rm tidal}&=&\frac{32}{5}\frac{c^5}{G}\eta^2x^{5}\left[    \left(\frac{18}{\chi_1}-12\right)\frac{c^{10}}{G^4}\frac{\lambda_1}{M^5}x^{5}-\left(\frac{176}{7 \chi _1}+\frac{1803}{28}-\frac{643}{4}\chi _1+\frac{155}{2}\chi _1^2\right)\frac{c^{10}}{G^4}\frac{\lambda_1}{M^5}x^{6} + (1\longleftrightarrow 2)    \right].
\end{eqnarray}
\end{widetext}
The total mass is $M=m_1+m_2$, where $m_1$ and $m_2$ are the component masses, $\eta=m_1m_2/M^2$ is the symmetric mass ratio, $x=(\pi G M f_{\rm gw} / c^3)^{2/3}$ is the PN expansion parameter, $f_{\rm gw}=2f_{\rm orb}$ is the GW frequency, $f_{\rm orb}$ is the binary's orbital frequency, and $\chi_1=m_1/M$ and $\chi_2=m_2/M$ are the two mass fractions.  Note that the PN order is labelled by the exponent on $x$ inside the square brackets, which is why these terms are referred to as 5PN and 6PN corrections.  Since the 5PN and 6PN tidal correction coefficients multiply $x^5$ and $x^6$ respectively, these effects will be insignificant at low frequencies and increasingly more significant at higher frequencies ($x\sim f_{\rm orb}^{2/3}$), as anticipated.  Appendix~\ref{PN_derivations} derives each tidally corrected PN waveform family from Eqs.~\eqref{CBC_dE} and~\eqref{CBC_dEdot}.  

The point-particle energy and luminosity are only known to 3.5PN order \cite{BIOPS}.  However, we add tidal corrections to the energy and luminosity that appear at 5PN and 6PN orders without knowing the higher order point-particle terms.  The justification for including the tidal corrections has typically been that they are always associated with the large coefficient $G\lambda_A [c^{2}/(G m_A)]^5\sim[c^2R_A/(G m_A)]^5\sim10^5$ \cite{Flanagan_Hinderer}.  Therefore, although they appear at high PN orders, the effect of the tidal terms on the binary's orbit are comparable to the effects of the 3PN and 3.5PN point-particle terms.  However, this claim was contradicted in \cite{Yagi_Yunes} because the tidal corrections are actually associated with the coefficient $[c^2R/(G M)]^5\sim10^3\ll[c^2R_A/(G m_A)]^5$, which is apparent from the form of Eqs.~\eqref{CBC_dE} and \eqref{CBC_dEdot}.  We show in Sec.~\ref{syst_error} that not knowing the higher order PN point-particle terms leads to significant systematic error when recovering tidal parameters.  Yagi and Yunes in \cite{Yagi_Yunes} and Favata in \cite{Favata} also discuss the importance of these unknown point-particle terms.

\subsection{Reparameterization of tidal parameters}
\label{define_LamT}

It becomes convenient to reparameterize the tidal parameters $(\lambda_1,\lambda_2)$ in terms of purely dimensionless parameters, which we call $(\tilde{\Lambda},\delta\tilde{\Lambda})$ \cite{Favata}.  Inspired by the $\tilde{\lambda}$ from \cite{Flanagan_Hinderer}, $\tilde{\Lambda}=32 G \tilde{\lambda} [c^2/(G M)]^5$ is essentially the entire 5PN tidal correction in all of the PN waveform families, while the 6PN tidal correction is a linear combination of $\tilde{\Lambda}$ and $\delta\tilde{\Lambda}$.  For example, the tidal corrections to the TaylorF2 phase later derived in Eq.~\eqref{F2_phase} of Appendix~\ref{PN_derivations} can equivalently be expressed as follows:
\begin{widetext}
\begin{equation}
\delta\psi_{\rm tidal}=\frac{3}{128\eta x^{5/2}}\left[ \left(-\frac{39}{2}\tilde{\Lambda}\right)x^5+ \left(-\frac{3115}{64}\tilde{\Lambda}+\frac{6595}{364}\sqrt{1-4\eta}\mbox{ }\delta\tilde{\Lambda}\right)x^6 \right],
\end{equation}
where
\begin{eqnarray}
\label{LT}
\tilde{\Lambda}&=&\frac{8}{13}\left[\left(1+7\eta-31\eta^2\right)\left(\Lambda_1+\Lambda_2\right)+\sqrt{1-4\eta}\left(1+9\eta-11\eta^2\right)\left(\Lambda_1-\Lambda_2\right)\right]\\
\label{dLT}
\delta\tilde{\Lambda}&=&\frac{1}{2}\left[\sqrt{1-4\eta}\left(1-\frac{13272}{1319}\eta+\frac{8944}{1319}\eta^2\right)\left(\Lambda_1+\Lambda_2\right) + \left(1-\frac{15910}{1319}\eta+\frac{32850}{1319}\eta^2+\frac{3380}{1319}\eta^3\right)\left(\Lambda_1-\Lambda_2\right)\right].
\end{eqnarray}
\end{widetext}
The dimensionless parameters $\Lambda_1=G\lambda_1[c^2/(G m_1)]^5$ and $\Lambda_2=G\lambda_2[c^2/(G m_2)]^5$, and we have assumed that $m_1>m_2$.  Though we choose to express $\tilde\Lambda$ and $\delta\tilde\Lambda$ in terms of dimensionless parameters as in Eqs.~\eqref{LT} and~\eqref{dLT}, they can be equivalently expressed more compactly in terms of dimensionful parameters, as can be inferred from Eq.~\eqref{F2_phase}. The parameters $(\tilde{\Lambda},\delta\tilde{\Lambda})$ were chosen such that they have the following convenient properties:
\begin{eqnarray}
\label{LT_prop}
\tilde{\Lambda}(\eta=1/4,\Lambda_1=\Lambda_2=\Lambda)&=&\Lambda\\
\label{dLT_prop}
\delta\tilde{\Lambda}(\eta=1/4,\Lambda_1=\Lambda_2=\Lambda)&=&0.
\end{eqnarray}
Setting $\eta=1/4$ implies that $m_1=m_2$.  Since all NSs have the same EOS, NSs with the same mass will also have the same value for $\Lambda$.  We have over-specified Eqs.~\eqref{LT_prop} and~\eqref{dLT_prop} for clarity.  We refer to $\tilde{\Lambda}$ as the {\it tidal deformability} of a BNS system throughout this work.  For more details regarding this reparameterization, see \cite{Favata}.\footnote{Note that, relative to \cite{Favata}, we have pulled out a factor of $\sqrt{1-4\eta}$ from our definition of $\delta\tilde\Lambda$ to allow for nonzero values of $\delta\tilde\Lambda$ when $\eta=1/4$.  This distinction enables the MCMC algorithm to fully explore the $\delta\tilde\Lambda$ parameter space even for equal mass systems.}


\section{Measurability of Tidal Influence}
\label{Sec_PE}

In this work, we use \texttt{lalinference\char`_mcmc} to run full Bayesian simulations for our parameter estimation investigation into the measurability of tidal deformability.  \texttt{lalinference\char`_mcmc} uses an MCMC sampling algorithm to calculate the posterior probability density function (PDF) of a detected CBC signal.  The algorithm is designed to efficiently explore a multi-dimensional parameter space in such a way that the density of parameter samples is a good approximation to the underlying posterior distribution.  In this section, we briefly outline the algorithm used by \texttt{lalinference\char`_mcmc}.  For a more comprehensive overview, we refer the reader to Refs. \cite{MCMC1,MCMC2,S6PE}.

\subsection{MCMC overview}
\label{MCMC_overview}

A true GW signal will be buried in detector noise.  Given a GW detection, the data stream segment $d(t)$ will have the following form in the time-domain:
\begin{equation}
\label{d=h+n}
d(t)=h(t)+n(t).
\end{equation}
The detector noise is denoted $n(t)$ while the pure GW signal is denoted $h(t)$.  Since no GWs have yet been detected by ground-based interferometers, our studies require simulated signals.  It is therefore customary to inject a modeled signal with chosen parameters into synthetic noise.

To determine the physical properties of a CBC system, we seek to map out the functional form of the posterior probability distribution ({\it posterior} for short) of its parameters.  Bayes' theorem relates the posterior $p(\vec{\theta}|d,m)$ for a set of parameters $\vec{\theta}$ given a model $m$ and data stream segment $d(t)$ to the prior probability distribution ({\it prior} for short) and the likelihood $p(d|\vec{\theta},m)$:
\begin{eqnarray}
\label{post}
p(\vec{\theta}|d,m)&=&\frac{p(\vec{\theta}|m)p(d|\vec{\theta},m)}{p(d|m)}\\
\label{unN_post}
&\propto&p(\vec{\theta}|m){\cal L}(d|\vec{\theta},m).
\end{eqnarray}
The notation $p(a|b)$ means the probability density of $a$ given $b$.  The posterior is the probability that the GW source modeled by $m$ that produced the data stream segment $d(t)$ has the physical properties $\vec{\theta}$.  The prior $p(\vec{\theta}|m)$ is the \emph{a priori} probability that the system modeled by $m$ has the physical properties $\vec{\theta}$.  The prior reflects everything that we know about the physical properties of any CBC system before attempting to determine the parameters of a specific source.  The evidence $p(d|m)$ is the probability of observing the data stream segment $d(t)$ with the model $m$.  The evidence is a normalization factor that can be used to compare how well different models would produce the data.  The likelihood ${\cal L}(d|\vec{\theta},m)=p(d|\vec{\theta},m)$ is the probability of observing the data stream segment $d(t)$ assuming the system that produced it is modeled by $m$ and has the physical properties $\vec{\theta}$.  The likelihood is a measure of how well the model $m$ with parameters $\vec{\theta}$ matches the data stream segment $d(t)$.  Assuming the noise is stationary and Gaussian, the functional form of the likelihood when using a single detector is \cite{Likelihood1,300yrs}
\begin{equation}
\label{unN_lik}
{\cal L}_{\rm det}(d|\vec{\theta},m)\propto\exp\left[-2\int_0^\infty \frac{\left| \tilde{d}_{\rm det}(f) - \tilde{m}(f,\vec{\theta}) \right|^2}{S_{\rm det}(f)}df \right].
\end{equation}
$S_{\rm det}(f)$ is the one-sided noise power spectral density (PSD), $\tilde d_{\rm det}(f)$ is the Fourier transform of the detector data stream segment, and $\tilde m(f,\vec{\theta})$ is a frequency-domain model for the waveform.  When using a network of GW detectors, the posterior probability becomes
\begin{equation}
p(\vec{\theta}|d,m)\propto p(\vec{\theta}|m)\prod_{\rm det}{\cal L}_{\rm det}(d|\vec{\theta},m).
\end{equation}

The MCMC algorithm used draws samples from the underlying posterior distribution $p(\vec{\theta}|d,m)$.  The samples can be binned to produce a histogram of the full multi-dimensional posterior distribution.  Posterior PDFs of fewer dimensions can be produced by marginalizing the full posterior over parameters that are not of interest.  For example, a 1D PDF for the tidal deformability $\tilde{\Lambda}$ can be found by integrating the posterior over all the other parameters:
\begin{equation}
p(\tilde{\Lambda}|d,m)=\int_{\vec{\theta}_{\rm other}} p(\vec{\theta}|d,m) d{\vec{\theta}_{\rm other}},
\end{equation}
where $\vec{\theta}_{\rm other}$ are all the parameters in the set $\vec{\theta}$ except $\tilde{\Lambda}$.  However, since the MCMC samples follow the posterior distribution, this integral is easily solved by simply binning only the parameters of interest (in this case $\tilde\Lambda$).

Various aspects of this algorithm have been fine-tuned to optimize speed and robustness and will be outlined in an upcoming methods paper.  This section is meant to merely provide an adequate overview of the parameter estimation pipeline used in this work.  We refer the interested reader to the following sources for more details \cite{MCMC1,MCMC2,S6PE}.

\subsection{Models, Parameters, and Priors}

Eq.~\eqref{unN_post} is used to calculate the posterior $p(\vec{\theta}|d,m)$, which is the quantity of interest, from the prior $p(\vec{\theta}|m)$ and likelihood ${\cal L}(d|\vec{\theta},m)$.  It depends on a model $m$, the model source parameters $\vec{\theta}$, and the prior distribution of each parameter.  The waveform models used in this work are the following tidally corrected PN waveform families, which we outline in Appendix~\ref{PN_derivations}: TaylorT1, TaylorT2, TaylorT3, TaylorT4, and TaylorF2.  To focus on purely EOS effects, we consider non-spinning BNS systems with no amplitude corrections.  (Parameter estimation can be just as easily performed with spinning waveforms, though slightly larger uncertainties in $\tilde\Lambda$ may arise for NSs with significant spins.)  These assumptions lead to the following 11-dimensional parameter space:
\begin{equation}
\vec{\theta}=\{{\cal M}_{\rm c},q,\tilde{\Lambda},\delta\tilde{\Lambda},D,\iota,\alpha,\delta,\phi_{\rm ref},t_{\rm ref},\psi\}.
\end{equation}
These parameters are: the chirp mass ${\cal M}_{\rm c}=\eta^{3/5}M$, the mass ratio $q=m_2/m_1$ where $m_1>m_2$, the distance to the binary $D$, the angle between the line of sight and the orbital axis $\iota$, the right ascension and declination of the binary $\alpha$ and $\delta$, the GW's polarization angle $\psi$, and the arbitrary reference phase and time $\phi_{\rm ref}$ and $t_{\rm ref}$.  Since $\Lambda_1$ and $\Lambda_2$ are highly correlated, we choose to parameterize in terms of $\tilde{\Lambda}$ and $\delta\tilde{\Lambda}$.  It is known that $\tilde{\Lambda}$ is comparatively more measurable than $\Lambda_1$ and $\Lambda_2$ individually \cite{Flanagan_Hinderer,Hinderer_Lackey}.  We use a uniform prior distribution in component masses between $1\mbox{ M}_\odot<m_{1,2}<30\mbox{ M}_\odot$, a uniform prior distribution in volume to $D<200\mbox{ Mpc}$, an isotropic prior distribution in sky location ($\alpha,\delta$) and emission direction ($\phi_{\rm ref},\iota$), a uniform prior distribution in polarization angle $\psi$, and a uniform prior distribution in $t_{\rm ref}$ over the data stream segment.  We use a uniform prior distribution in $\tilde{\Lambda}$ between $0<\tilde{\Lambda}<3000$ and a uniform prior distribution in $\delta\tilde{\Lambda}$ between $-500<\delta\tilde{\Lambda}<500$.  These ranges were chosen to include effects from the majority of possible NS EOSs.\footnote{Note that $\tilde{\Lambda}$ may exceed 3000 for low mass NSs with a stiff EOS.  However, this upper bound does not affect the results in this paper, because the posterior is found to be fully contained within the region of prior support for all cases considered.}

Since we are concerned only with measuring EOS effects on BNS signals, we fixed all the injected signals to have the exact same sky position ($\alpha=0.648522$ and $\delta=0.5747465$), orientation ($\iota=0.7240786$), and polarization ($\psi=2.228162$) for comparison purposes.  We vary the strength of injected signals by adjusting $D$.  We also use a 3-detector advanced LIGO/Virgo network.  We use the PSD of the two advanced LIGO detectors under the zero-detuned high power configuration \cite{ZDHP} and the parameterized advanced Virgo PSD based on Eq.~6 of \cite{Virgo_PSD}.  Injection and template waveforms all have a low frequency cutoff at $f_{\rm low}=30$ Hz and end when the system reaches $f_{\rm high}=f_{\rm ISCO}=c^3/(6^{3/2}\pi G M)$, where $f_{\rm ISCO}$ is the GW frequency when the binary reaches the inner-most stable circular orbit (ISCO).  Tests using different high frequency cutoffs did not affect our results in any noticeable way.

\subsection{Measurability of Tidal Deformability}
\label{measure_tides}

In order to simulate the parameter estimation of a GW signal, one typically injects a model waveform into a data stream segment consisting of simulated detector noise.  The strength of the injected signal relative to the detector noise is characterized by the SNR.  The SNR $\rho_{\rm det}$ of an injection into a single GW detector is
\begin{equation}
\rho_{\rm det}=\sqrt{4\int_0^\infty \frac{|\tilde{m}(f,\vec{\theta})|^2}{S_{\rm det}(f)}df},
\end{equation}
where $\tilde{m}(f,\vec{\theta})$ is the injected waveform model in the frequency domain and $S_{\rm det}(f)\delta(f-f^\prime)/2=\langle\tilde{n}_{\rm det}{}^*(f^\prime)\tilde{n}_{\rm det}(f)\rangle$, where $\tilde{n}_{\rm det}(f)$ is the Fourier transform of the detector's noise.  For a collection of detectors, the network SNR $\rho_{\rm net}$ is defined to be
\begin{equation}
\rho_{\rm net}=\sqrt{\sum_{\rm det} \rho_{\rm det}^2}.
\end{equation}

We report on the optimal measurability of tidal influences in BNS systems assuming a 3-detector LIGO/Virgo network.  We follow a similar procedure to the one used in \cite{Carl_Ben_PE}, which details the statistical uncertainties in the mass parameters and sky location parameters of BNS systems that are expected to be achieved with advanced detectors.  While one typically injects a signal into synthetic noise, we sometimes choose not to add synthetic noise to our injected signal, which essentially means that we set $n(t)=0$ in Eq.~\eqref{d=h+n}.  However, we still calculate the likelihood and the network SNR by dividing by the detector PSD, which is the variance of the noise.  In this way, we incorporate the overall effect of noise without dealing with the statistical fluctuations of individual noise realizations.  We refer to this procedure as ``injecting into zero-noise'' \cite{Carl_Ben_PE}.

We inject into zero-noise for two reasons.  The first reason is to report measurement uncertainties for typical systems.  However, individual results depend on individual realizations of the noise at the time of detection.  It is shown in \cite{0noise} that the average posterior PDF, or the posterior distribution averaged over noise realizations, is recovered by setting the noise to zero.  We can therefore get reliable estimates for the mean measurement uncertainty of tidal parameters recovered from a signal injected into many different noise realizations by simply injecting that signal into zero-noise \cite{0noise, Carl_Ben_PE, Carl_FM, Sampson}.  This saves us from having to perform many MCMC simulations with different noise realizations.  While this approach only considers the overall effect of noise, we discuss the effect of individual noise realizations in Sec.~\ref{stat_error}.  The second reason for injecting into zero-noise, which we use in Sec.~\ref{syst_error}, is to isolate the effects of systematic error in our analysis.  By injecting into zero-noise, we are able to disentangle modeling bias from noise realization effects without having to perform many MCMC simulations, which are computationally expensive \cite{Littenberg}.

\begin{figure*}[t]
	\begin{minipage}[b]{0.3292\textwidth}
	\centering
	\includegraphics[width=\textwidth]{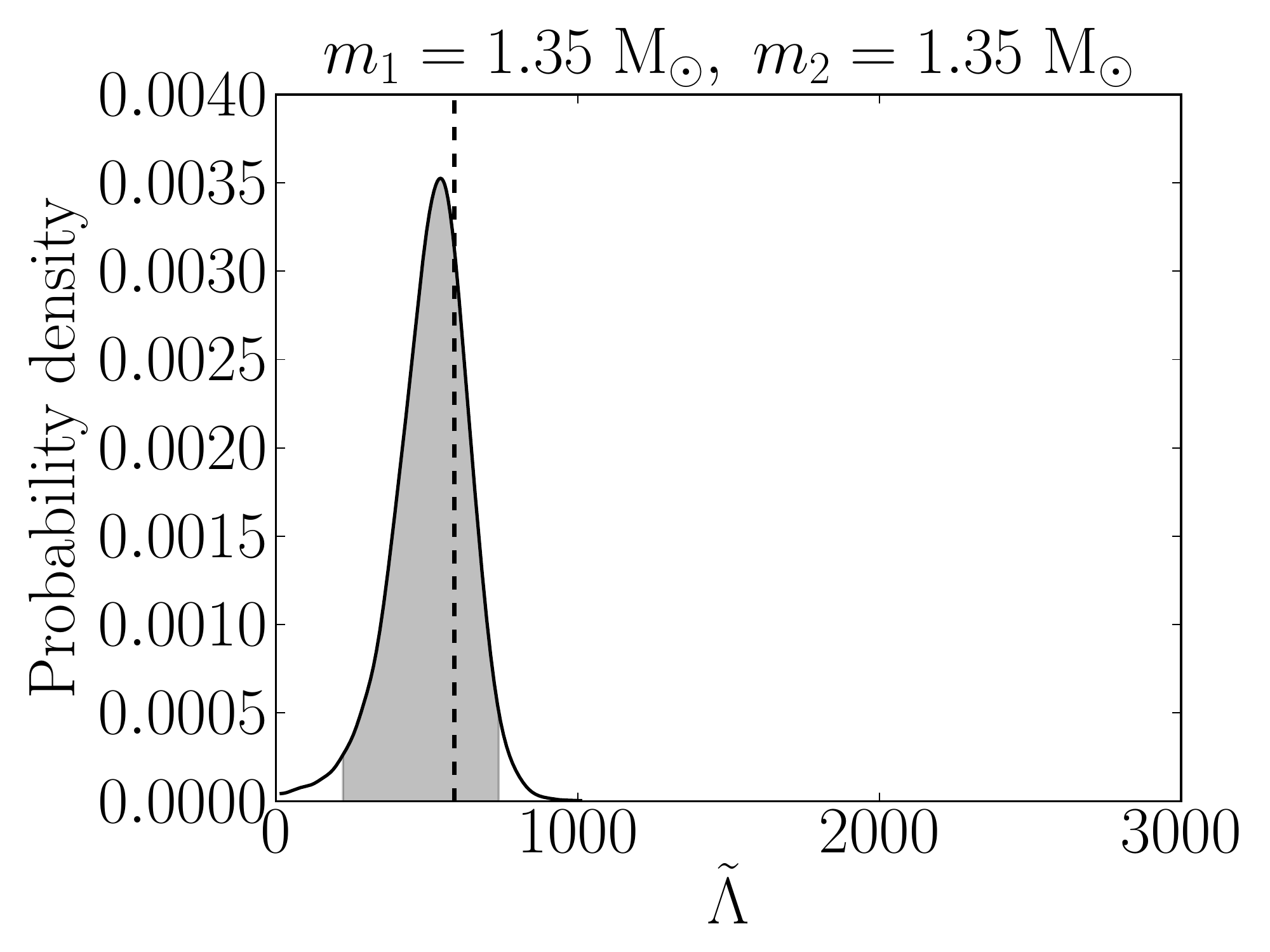}
	\end{minipage}
	\begin{minipage}[b]{0.3292\textwidth}
	\centering
	\includegraphics[width=\textwidth]{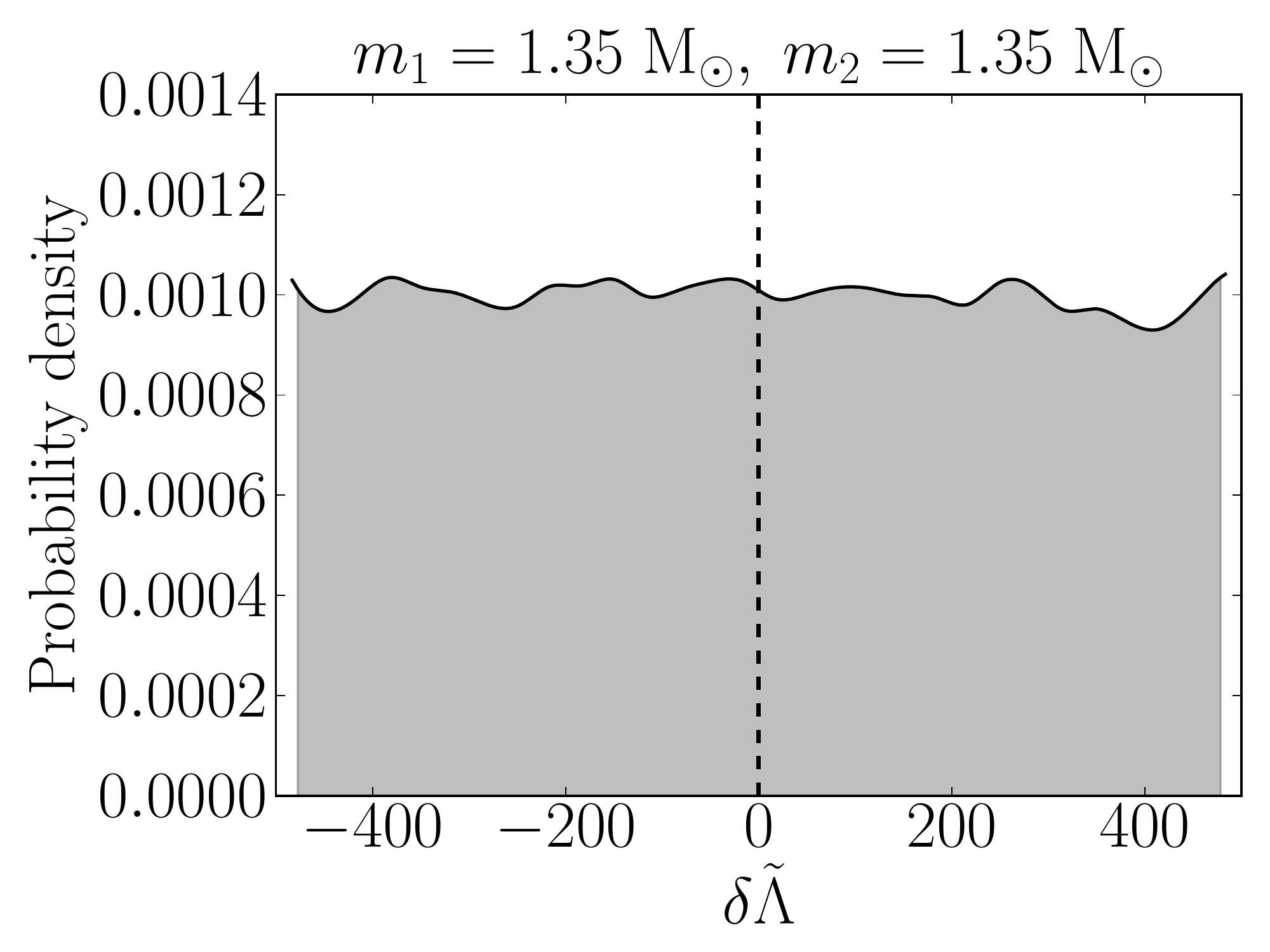}
	\end{minipage}
	\begin{minipage}[b]{0.3292\textwidth}
	\centering
	\includegraphics[width=\textwidth]{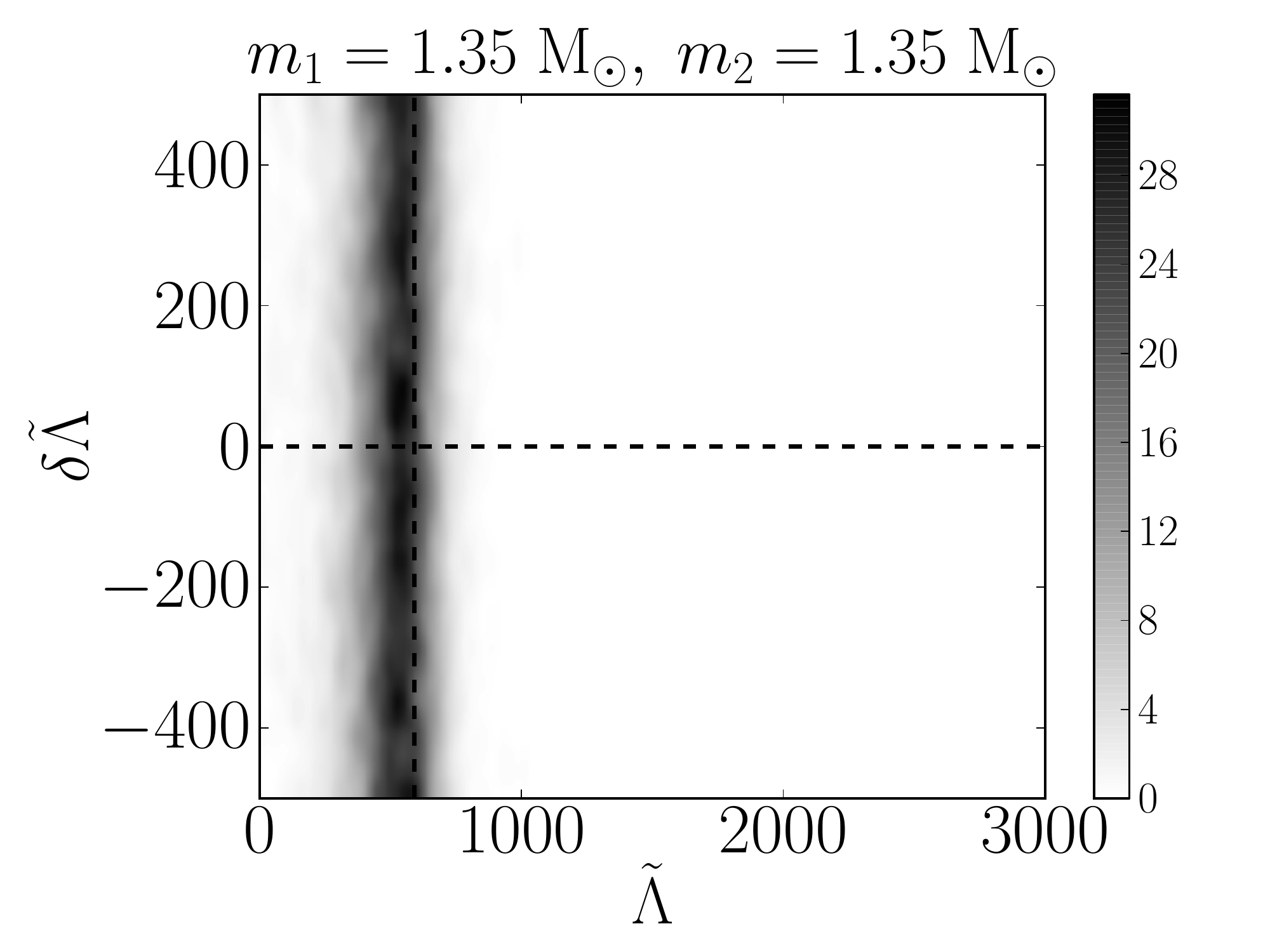}
	\end{minipage}
\caption{\label{lam_LamT}{\footnotesize Marginalized 1D (left and middle) and 2D (right) posterior probability density functions for $\tilde\Lambda$ and $\delta\tilde\Lambda$ of a 1.35 M$_\odot$:1.35 M$_\odot$ BNS system with $\rho_{\rm net}=32.4$.  The shaded regions in the 1D PDFs enclose 2$\sigma$ (95\%) confidence regions.  The color bar in the 2D PDF labels the (unnormalized) probability density.  The injected values for $\tilde\Lambda$ and $\delta\tilde\Lambda$ are consistent with the MPA1 EOS model \cite{4PieceFit} and are marked with straight dashed lines.  These plots are PDFs smoothed with a Gaussian kernel density estimator.  For these results, we injected into zero-noise (see Sec.~\ref{measure_tides}).}}
\end{figure*}

In Fig.~\ref{lam_LamT}, we present the 1D and 2D posterior PDFs for $\tilde{\Lambda}$ and $\delta\tilde{\Lambda}$ of a typical BNS system.  The true signal was injected with $\rho_{\rm net}=32.4$, which is considered very large (perhaps a one-per-year event by 2019 \cite{ObservingScenarios}).  We use tidally corrected TaylorF2 waveforms for the injected waveform as well as for the recovery template waveforms.  The injection has the following properties: $m_1=m_2=1.35\mbox{ M}_\odot$, $\tilde{\Lambda}=590.944$, and $\delta\tilde{\Lambda}=0$, which is consistent with the MPA1 EOS model\footnote{We actually use the parameterized EOS presented in \cite{4PieceFit} that matches the theoretical MPA1 EOS, as well as many other theoretical EOSs, to a few percent.  This approximation is used throughout this work for our convenience.  Since the EOS is only used to estimate injected $\tilde\Lambda$ values, our results will not be affected by this approximation.} \cite{4PieceFit}.  We find that the injected value of $\tilde{\Lambda}$ is well recovered.\footnote{The peak of the 1D PDF for $\tilde\Lambda$ is consistently found to be displaced from the injected value for equal mass and near equal mass systems.  This is a result of marginalizing over the other ten parameters \cite{Carl_Ben_PE}, in particular the mass ratio $q$, whose prior distribution caps off at $q=m_2/m_1=1$.}  However, advanced detectors are not able to discern $\delta\tilde{\Lambda}$ contributions to the waveform even at a network SNR of 32.4.  This is expected because $\delta\tilde\Lambda$ only shows up in the 6PN tidal correction, which is ${\cal O}$(10\%) as big as the 5PN term, and additionally contributes little to the 6PN correction since $\delta\tilde\Lambda/\tilde\Lambda\sim0$--0.01 \cite{Favata}.

In Table~\ref{Tab_measurability} we outline the measurement uncertainties for the tidal deformability parameter $\tilde{\Lambda}$ for several equal mass and unequal mass BNS systems.  We compute the 1$\sigma$ and 2$\sigma$ measurement uncertainty interval by determining the smallest interval in $\tilde\Lambda$ that contains 68\% and 95\% of the total marginalized posterior probability.  We then report the lower and upper bound on this confidence interval.  The $1\sigma$ confidence interval for a 1.35 M$_\odot$:1.35 M$_\odot$ BNS system consistent with the MPA1 EOS model is (382.0,636.7) for $\rho_{\rm net}=30$.  We find that the measurability of the other parameters are not noticeably affected by including tidal parameters in our analysis.

\begin{table*}[t]
	\caption{{\footnotesize The 1$\sigma$ (68\%) and 2$\sigma$ (95\%) confidence intervals (min,max) for $\tilde\Lambda$.  The BNS systems considered are labeled by their injected masses and tidal deformability $\tilde\Lambda$.  Both equal mass and unequal mass systems ranging from $m_{\rm min}=1.20$ M$_\odot$ to $m_{\rm max}=2.10$ M$_\odot$ are considered.  The injected values for $\tilde\Lambda$ are consistent with the MPA1 EOS model \cite{4PieceFit}.  We report confidence intervals for systems with a network SNR of both 20 and 30.  For these results, we injected into zero-noise (see Section~\ref{measure_tides}).}}
\begin{ruledtabular}
		\begin{tabular}{%
			D{.}{.}{-1}@{\extracolsep{\fill}}%
			D{.}{.}{-1}@{\extracolsep{\fill}}%
			D{.}{.}{-1}@{\extracolsep{\fill}}%
			c@{\extracolsep{0pt}( }%
			D{.}{.}{2}@{\extracolsep{0pt} , }%
			D{.}{.}{1}@{\extracolsep{\fill} )}%
			c@{\extracolsep{0pt}( }%
			D{.}{.}{2}@{\extracolsep{0pt} , }%
			D{.}{.}{1}@{\extracolsep{\fill} )}%
			c@{\extracolsep{0pt}( }%
			D{.}{.}{2}@{\extracolsep{0pt} , }%
			D{.}{.}{1}@{\extracolsep{\fill} )}%
			c@{\extracolsep{0pt}( }%
			D{.}{.}{2}@{\extracolsep{0pt} , }%
			D{.}{.}{1}@{\extracolsep{\fill} )}}
		\multicolumn{4}{c}{}
		& \multicolumn{5}{c}{$\rho_{\rm net}=20$}
		& \multicolumn{1}{c}{}
		& \multicolumn{5}{c}{$\rho_{\rm net}=30$}
		\\
		\cline{4-9}\cline{10-15}
		\multicolumn{1}{c}{$m_1$ (M$_\odot$)}
		& \multicolumn{1}{c}{$m_2$ (M$_\odot$)}
		& \multicolumn{1}{c}{$\tilde{\Lambda}$}
		& \multicolumn{1}{c}{}
		& \multicolumn{2}{c}{1$\sigma$} 
		& \multicolumn{1}{c}{}
		& \multicolumn{2}{c}{2$\sigma$}
		& \multicolumn{1}{c}{}
		& \multicolumn{2}{c}{1$\sigma$} 
		& \multicolumn{1}{c}{}
		& \multicolumn{2}{c}{2$\sigma$}
		\\
		\hline
		1.20  & 1.20  & 1135.630 && 553.8 & 1258.1  && 134.6 & 1700.1  && 838.7 & 1193.8  && 516.6 & 1359.4 \\
		1.35  & 1.35  & 590.944 && 251.3 & 690.2 && 60.7 & 963.0 && 382.0 & 636.7 && 182.3 & 750.8 \\
		1.50  & 1.50  & 318.786 && 113.2 & 398.9 && 22.9 & 576.8 && 162.1 & 357.4 && 63.9 & 447.7 \\
		1.65  & 1.65  & 175.963 && 54.5 & 250.2 && 9.6 & 377.2 && 63.5 & 213.9 && 14.0 & 290.8 \\
		1.80  & 1.80  & 98.191 && 29.2 & 176.8 && 4.9 & 274.9 && 28.9 & 136.1 && 5.0 & 196.8 \\
		1.95  & 1.95  & 54.670 && 20.1 & 132.5 && 3.5 & 214.4 && 16.6 & 96.1 && 2.6 & 148.2 \\
		2.10  & 2.10  & 29.844 && 14.8 & 104.8 && 2.1 & 174.4 && 11.7 & 74.0 && 1.9 & 118.6 \\
		1.35  & 1.20  & 820.610 && 433.7 & 1017.6  && 102.7 & 1381.7  && 612.9 & 941.3 && 340.7 & 1094.6 \\
		1.35  & 1.50  & 435.585 && 200.0 & 574.9 && 44.4 & 814.5 && 282.5 & 518.0 && 125.5 & 626.1 \\
		1.35  & 1.65  & 328.177 && 196.1 & 570.5 && 45.5 & 834.6 && 221.3 & 495.9 && 85.5 & 619.1 \\
		1.35  & 1.80  & 252.398 && 155.1 & 593.1 && 33.0 & 907.0 && 155.9 & 433.5 && 45.5 & 598.6 \\
		1.35  & 1.95  & 197.899 && 119.0 & 546.9 && 21.5 & 922.6 && 107.3 & 348.2 && 24.7 & 489.1 \\
		1.35  & 2.10  & 157.974 && 90.7 & 445.4 && 15.8 & 819.9 && 79.3 & 296.8 && 16.2 & 424.9 \\
		\end{tabular}
	\label{Tab_measurability}
\end{ruledtabular}
\end{table*}

We can also compare our MCMC results to a few FM results.  The FM study by Favata \cite{Favata} uses tidally corrected PN waveforms with a high frequency cutoff of 1000 Hz.  Favata finds the 1$\sigma$ measurement uncertainty of the tidal deformability parameter to be roughly 27\% for a 1.40 M$_\odot$:1.40 M$_\odot$ BNS system with $\tilde\Lambda\approx600$ at an SNR of 30.  Damour, Nagar, and Villain \cite{Damour_EOB} use tidally corrected EOB waveforms that end at contact.  In their FM study, they find a slightly better measurement uncertainty of roughly 21\% for a 1.40 M$_\odot$:1.40 M$_\odot$ BNS system with $\tilde\Lambda\approx600$ at an SNR of 30.\footnote{Since \cite{Damour_EOB} does not include the measurement uncertainty of a BNS system with $\tilde\Lambda\approx600$, this measurement uncertainty was estimated via interpolation.}  This improvement is likely due to the extra high frequency information included in the EOB waveforms.  Read {\it et al.}\ \cite{Read_ME} use NR waveforms in their FM study, though they rely on a somewhat crude finite difference approximation.  For a 1.35 M$_\odot$:1.35 M$_\odot$ BNS system with $\tilde\Lambda\approx600$ at an SNR of 30, they find a measurement uncertainty of roughly 16\% with full hybrid waveforms, though they do not consider correlations with other parameters.\footnote{The finite difference approximation is between the EOS H and HB: $\sigma_{\tilde\Lambda}=(\tilde\Lambda_{\rm H}-\tilde\Lambda_{\rm HB})/||h_{\rm H}-h_{\rm HB}||=85$, which results in a measurement uncertainty of $\sigma_{\tilde\Lambda}/\tilde\Lambda_{\rm H}=0.16$ when scaled to an SNR of 30.}  Again, the slight increase in measurability is likely due to the additional high frequency information included in their waveforms.  In our MCMC study, we find the measurement uncertainty of the tidal deformability parameter to be roughly 21\% for a 1.35 M$_\odot$:1.35 M$_\odot$ BNS system with $\tilde\Lambda\approx600$ at an SNR of 30 in a single advanced LIGO detector.  This is in general agreement with existing FM calculations.


\section{Constraining NS EOS}
\label{Sec_NS_EOS}

The NS EOS describes the structure of all cold NSs in equilibrium by relating NS state variables, such as pressure and density.  Simultaneous NS mass-radius measurements, or equivalently mass-$\lambda$ measurements, can highly constrain the NS EOS \cite{Lindblom_MR,Lindblom_SpecI,Lindblom_SpecII}.   While many accurate NS mass measurements have been made, corresponding radius measurements are still needed \cite{NS_EOS_Review}.

While $\Lambda_1\sim (R_1/m_1)^5$ and $\Lambda_2\sim (R_2/m_2)^5$ are poorly measured by advanced GW detectors due to their strong correlation, the tidal deformability parameter $\tilde{\Lambda}$, which is a linear combination of $(\Lambda_1,\Lambda_2)$, is better measured.  Ground-based interferometers are most adept at measuring a system's chirp mass ${\cal M}_{\rm c}$.  In the same way that a binary's chirp mass is a mass-like parameter that contains information about the mass of both components, the fifth root of the tidal deformability parameter $\tilde\Lambda^{1/5}$ can be thought of as a dimensionless radius-like parameter that contains information about the radius of both components.  While GW detectors may not be able to simultaneously constrain the mass and radius of individual NS's, we show that they can simultaneously constrain the mass-like and radius-like parameters describing the binary system as a whole.  To further this analogy, we choose to define a conveniently scaled dimensionful radius-like parameter ${\cal R}_{\rm c}=2 G {\cal M}_{\rm c}\tilde\Lambda^{1/5}/c^2$, which we call the binary's {\it chirp radius}.  Therefore, making a ${\cal M}_{\rm c}$--${\cal R}_{\rm c}$ measurement of a CBC system is analogous to making a mass--radius measurement of a single NS star.  Note that the component masses and radii are entangled in the former case and are only determined in combination.  The question then becomes: ``Does measuring the chirp mass and the chirp radius as opposed to the individual mass and individual radius contain enough information to constrain the NS EOS?''

\begin{figure*}[t]
	\begin{minipage}[b]{0.4971\textwidth}
	\centering
	\includegraphics[width=\textwidth]{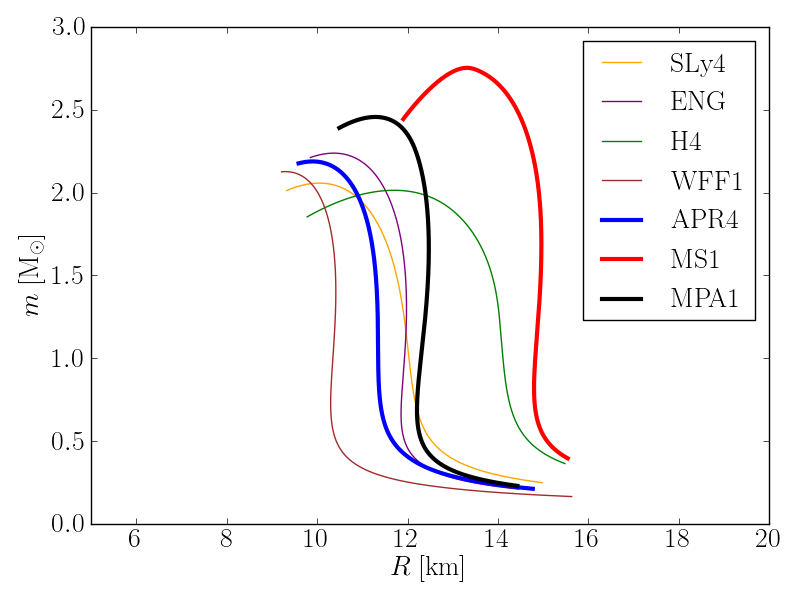}
	\end{minipage}
	\begin{minipage}[b]{0.4971\textwidth}
	\centering
	\includegraphics[width=\textwidth]{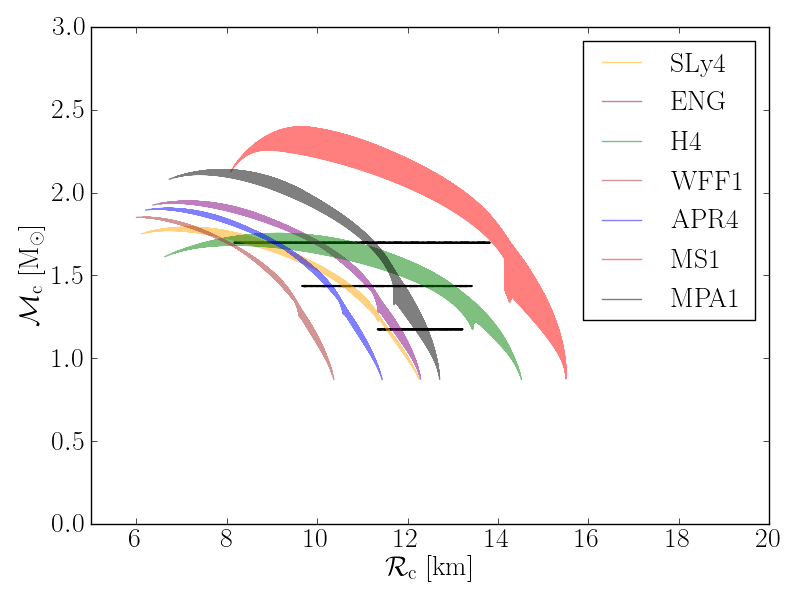}
	\end{minipage}
	\
	\begin{minipage}[b]{0.3292\textwidth}
	\centering
	\includegraphics[width=\textwidth]{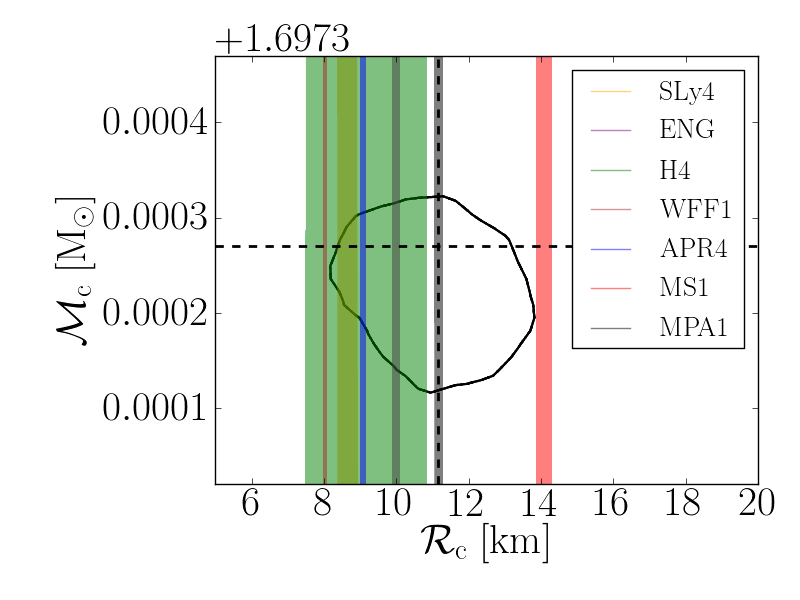}
	\end{minipage}
	\begin{minipage}[b]{0.3292\textwidth}
	\centering
	\includegraphics[width=\textwidth]{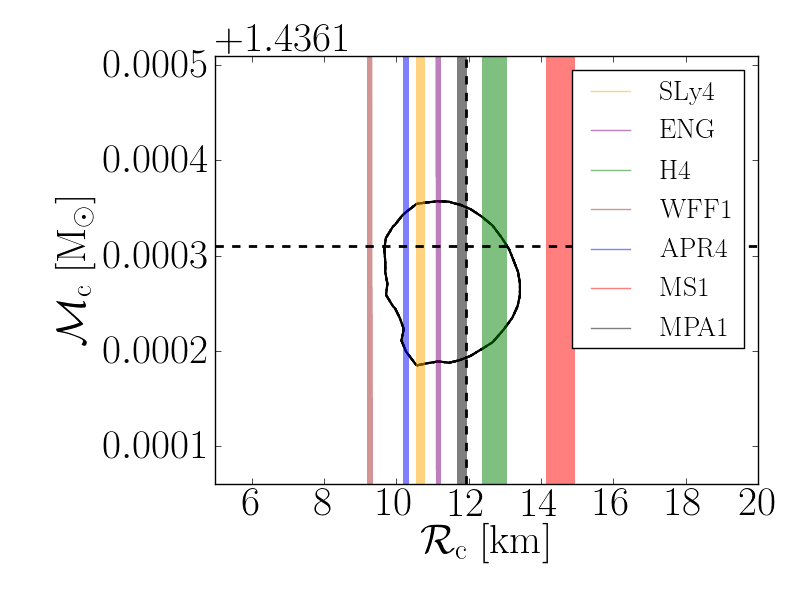}
	\end{minipage}
	\begin{minipage}[b]{0.3292\textwidth}
	\centering
	\includegraphics[width=\textwidth]{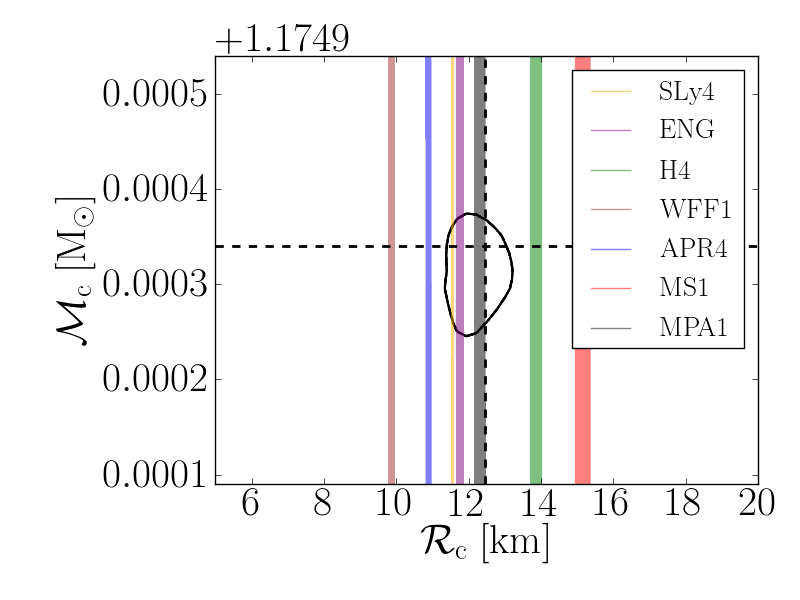}
	\end{minipage}
\caption{\label{EOS_plots}{\footnotesize NS mass-radius plot for a sample of NS EOS models found in the literature \cite{4PieceFit} (top left).  The ${\cal M}_{\rm c}$--${\cal R}_{\rm c}$ plot (top right), where ${\cal R}_{\rm c}$ is defined in Sec.~\ref{Sec_NS_EOS}, depicts the same EOSs as the mass-radius plot now smeared out due to the extra degrees of freedom from not specifying individual masses and radii.  We consider NSs with masses that range from 1 M$_\odot$ to the maximum allowed mass for each EOS.  The three horizontal, black lines are the $1\sigma$ (68\%) confidence regions of three recovered injections.  The three bottom plots are zoomed-in to show these recovered injections more clearly.  The injected values for ${\cal M}_{\rm c}$ and ${\cal R}_{\rm c}$ are consistent with the MPA1 EOS model and are marked with straight, dashed lines.  For these results, we injected into zero-noise (see Sec.~\ref{measure_tides}).}}
\end{figure*}

In Fig.~\ref{EOS_plots}, we take a mass-radius plot with multiple theoretical EOS curves \cite{4PieceFit} (upper left) and transform it into a ${\cal M}_{\rm c}$--${\cal R}_{\rm c}$ plot with the same EOS curves, now smeared out due to the extra degrees of freedom from not specifying individual masses and radii (upper right).  The three horizontal, black lines are the $1\sigma$ confidence regions of three recovered injections.  Because chirp mass is so well measured, these confidence regions appear to be lines due to the aspect ratio of this plot.  The three bottom plots in Fig.~\ref{EOS_plots} are zoomed-in plots of each recovered injection.  From left to right, the important parameters for each injection are: $m_1=m_2=1.50\mbox{ M}_{\odot}$ and $\tilde{\Lambda}=\Lambda_1=\Lambda_2=318.786$, $m_1=m_2=1.35\mbox{ M}_{\odot}$ and $\tilde{\Lambda}=\Lambda_1=\Lambda_2=590.944$, and $m_1=m_2=1.20\mbox{ M}_{\odot}$ and $\tilde{\Lambda}=\Lambda_1=\Lambda_2=1135.63$.  The injections all correspond to the EOS MPA1 \cite{4PieceFit} and have $\rho_{\rm net}=30$.  Fig.~\ref{EOS_plots} demonstrates that simultaneous ${\cal M}_{\rm c}$--${\cal R}_{\rm c}$ measurements can indeed constrain the NS EOS.  However, because certain regions of parameter space can be described by overlapping EOS curves, BNS observations with varying values for chirp mass will likely need to be observed before tight constraints on the NS EOS can be made with this approach.

This inversion of ${\cal M}_{\rm c}$--${\cal R}_{\rm c}$ measurements to EOS constraints is similar to the inverse stellar structure problem described in \cite{Lindblom_MR,Lindblom_SpecI,Lindblom_SpecII}.  Other methods for constraining the NS EOS with GW detectors are discussed in Sec.~\ref{Sec_Conclusion}.


\section{Sources of Error}
\label{Sec_Error}

Sources of error in estimating the parameters of a CBC system given its gravitational signal can be categorized as statistical and/or systematic.  Statistical error is due to the presence of random detector noise.  In Sec.~\ref{measure_tides}, we focused on the overall effect of detector noise.  In this section, our focus is on the effect of individual noise realizations.  The kind of systematic error that we are studying arises because our template waveforms only approximate true signals.  Statistical error is SNR-dependent, since it depends on the relative strength of the signal to the detector noise, while systematic error is SNR-independent.  In this section, we present the effects of both systematic error and individual noise realizations on the ability of advanced ground-based interferometers to measure tidal deformability.

\subsection{Systematic Error}
\label{syst_error}

The PN approximation to the energy and luminosity of a CBC system is an expansion of the equations of motion about small characteristic velocities, or small frequencies ($v\sim f_{\rm gw}^{1/3}$).  Currently, the point-particle corrections to the CBC energy and luminosity are known to 3.5PN order \cite{BIOPS}.  While PN waveforms match a true GW signal at small frequencies, they are unreliable at high frequencies.  Since tidal influences become significant at high frequencies, it is expected that the systematic error from having unreliable waveforms at high frequencies will bias the recovery of tidal parameters.  The question is: ``By how much?''

We expect that the deviation of PN waveform families away from the true CBC waveform will be comparable to the amount that they deviate away from each other.  All of the PN waveform families are accurate to the same PN order but differ from one another at higher orders.  We use the fact that we cannot say which PN family is more accurate as a simple way to parameterize our ignorance of unknown higher order PN terms.  We test systematic bias by injecting one PN waveform family and recovering with another.  Because all PN waveform families are considered viable, this gives at least a lower bound on the systematic error due to modeling bias.  In this way, we can get an order of magnitude estimate of the systematic bias that results from using waveforms that are unreliable at high frequencies to estimate tidal parameters whose effects arise at high frequencies.

\begin{figure*}[t]
	\begin{minipage}[b]{0.3292\textwidth}
	\centering
	\includegraphics[width=\textwidth]{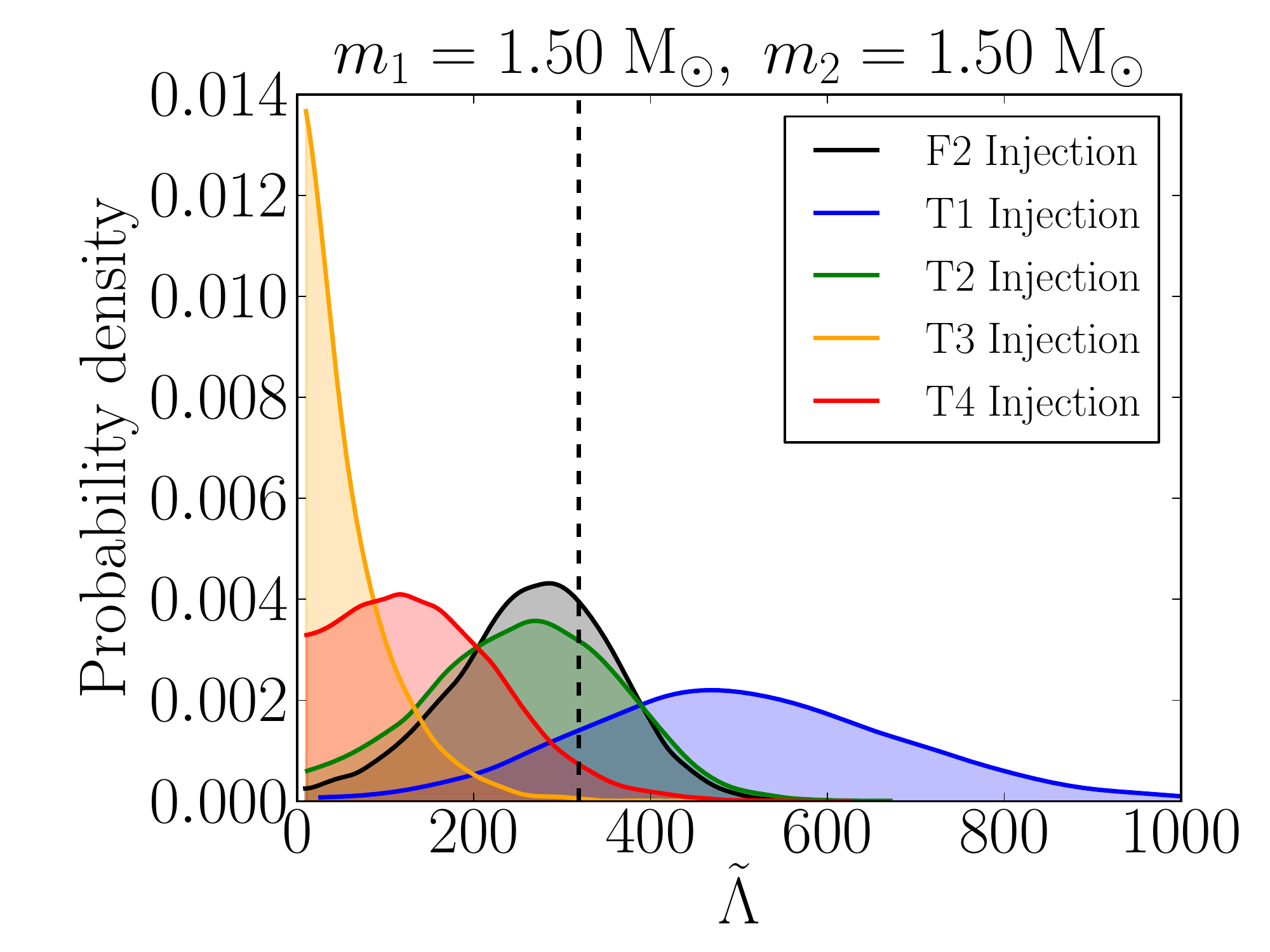}
	\end{minipage}
	\begin{minipage}[b]{0.3292\textwidth}
	\centering
	\includegraphics[width=\textwidth]{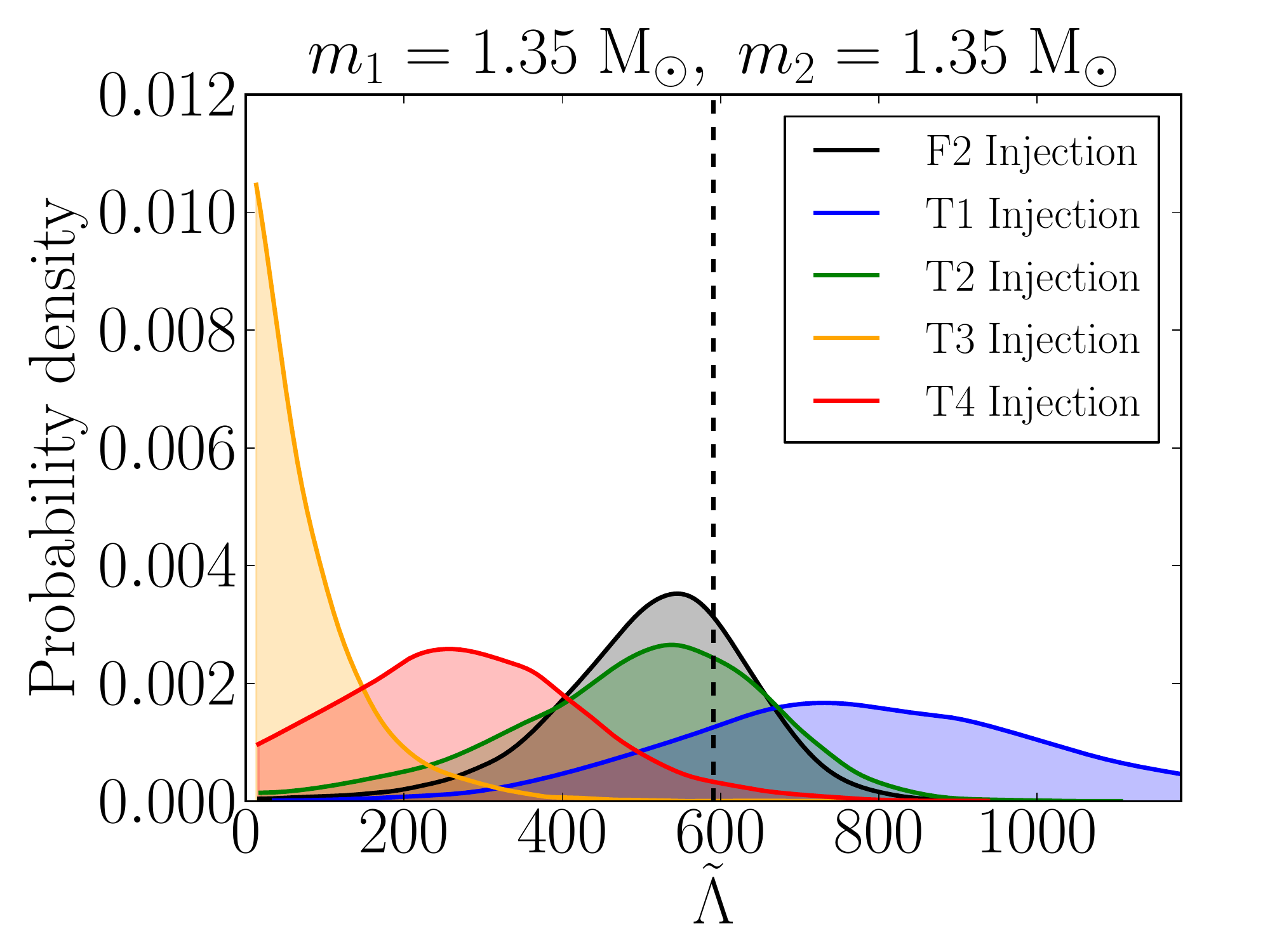}
	\end{minipage}
	\begin{minipage}[b]{0.3292\textwidth}
	\centering
	\includegraphics[width=\textwidth]{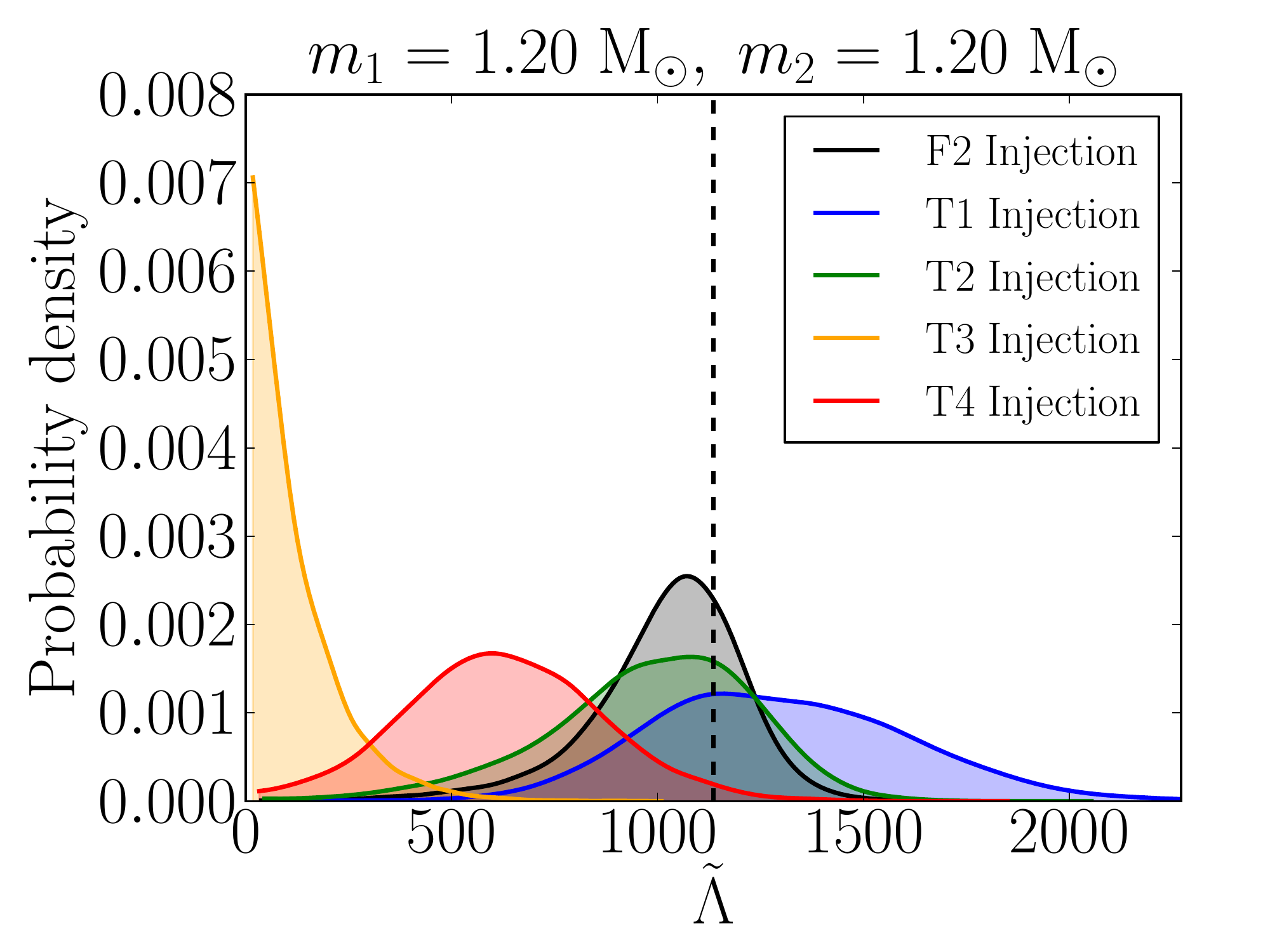}
	\end{minipage}
	\
\caption{\label{Sys_LamT}{\footnotesize Marginalized 1D posterior probability density functions for $\tilde\Lambda$ of three BNS systems (labelled by the masses in the title) each with $\rho_{\rm net}=32.4$.  The injected $\tilde\Lambda$ values are consistent with the MPA1 EOS model \cite{4PieceFit} and are marked with straight, dashed lines.  These plots are PDFs smoothed with a Gaussian kernel density estimator.  To generate a single plot, we inject BNS signals modeled by each of the five PN waveform families derived in Appendix~\ref{PN_derivations}.   Though the waveform family for each signal is different, the injected waveform parameters are identical.  The five PDFs, which are labelled by the injected waveform family, are all recovered using TaylorF2 waveform templates.  The deviation of each peak away from the injected value is due to the systematic error in the PN waveform approximants.  For these results, we injected into zero-noise (see Sec.~\ref{measure_tides}).}}
\end{figure*}

In Fig.~\ref{Sys_LamT}, we present example 1D posterior PDFs for $\tilde\Lambda$.  We inject signals from each of the five PN waveform families derived in Appendix~\ref{PN_derivations} but only recover with TaylorF2 templates.  Since injected waveforms are only generated once while template waveforms are generated millions of times during an MCMC run, we only use TaylorF2 templates, because they are generated much faster than the other PN waveform families.  The injected component masses are labeled in each figure's title, while the injected value of $\tilde\Lambda$, which is consistent with the EOS labeled in the legend, is marked by a dashed, vertical line.  Each injection has a network SNR of 32.4 and was injected into zero-noise in order to isolate systematic error from statistical error.  (Remember that the effects of noise are not completely ignored by injecting into zero-noise.  The PSD is still used to calculate likelihood and network SNR.)  While we only present three mass combinations and one EOS model in Fig.~\ref{Sys_LamT}, we also find similar results when considering several other equal and unequal mass combinations and EOS models.

We find that systematic error can be significant in each of the mass combinations and EOSs considered.  In particular, the TaylorT4 waveform family has been found to be remarkably similar to equal mass numerical relativity (NR) waveforms \cite{Radice_T4vsNR}.  Therefore, for a typical $m_1=m_2=1.35\mbox{ M}_{\odot}$ BNS system with a moderate EOS, say MPA1, systematic error will likely bias the maximum likelihood recovery of $\tilde\Lambda$ by $(\tilde\Lambda_{\rm inj}-\tilde\Lambda_{\rm rec})/\tilde\Lambda_{\rm inj}\sim$50\%.

It is also interesting to note that the TaylorT3 injected waveforms are all recovered with little to no tidal contribution with TaylorF2 templates.  Additionally, the TaylorT3 injected waveforms were recovered with a chirp mass bias of roughly twice its standard deviation, whereas none of the other injected waveforms were recovered with noticeable bias in chirp mass. It was previously seen in \cite{BIOPS} that the TaylorT3 approximant agrees poorly with other PN approximants due to its peculiar termination conditions, and we suspect this also explains the biases seen here.

\subsection{Noise Realizations}
\label{stat_error}

\begin{figure*}[t]
	\begin{minipage}[b]{0.31\textwidth}
	\centering
	\includegraphics[width=\textwidth]{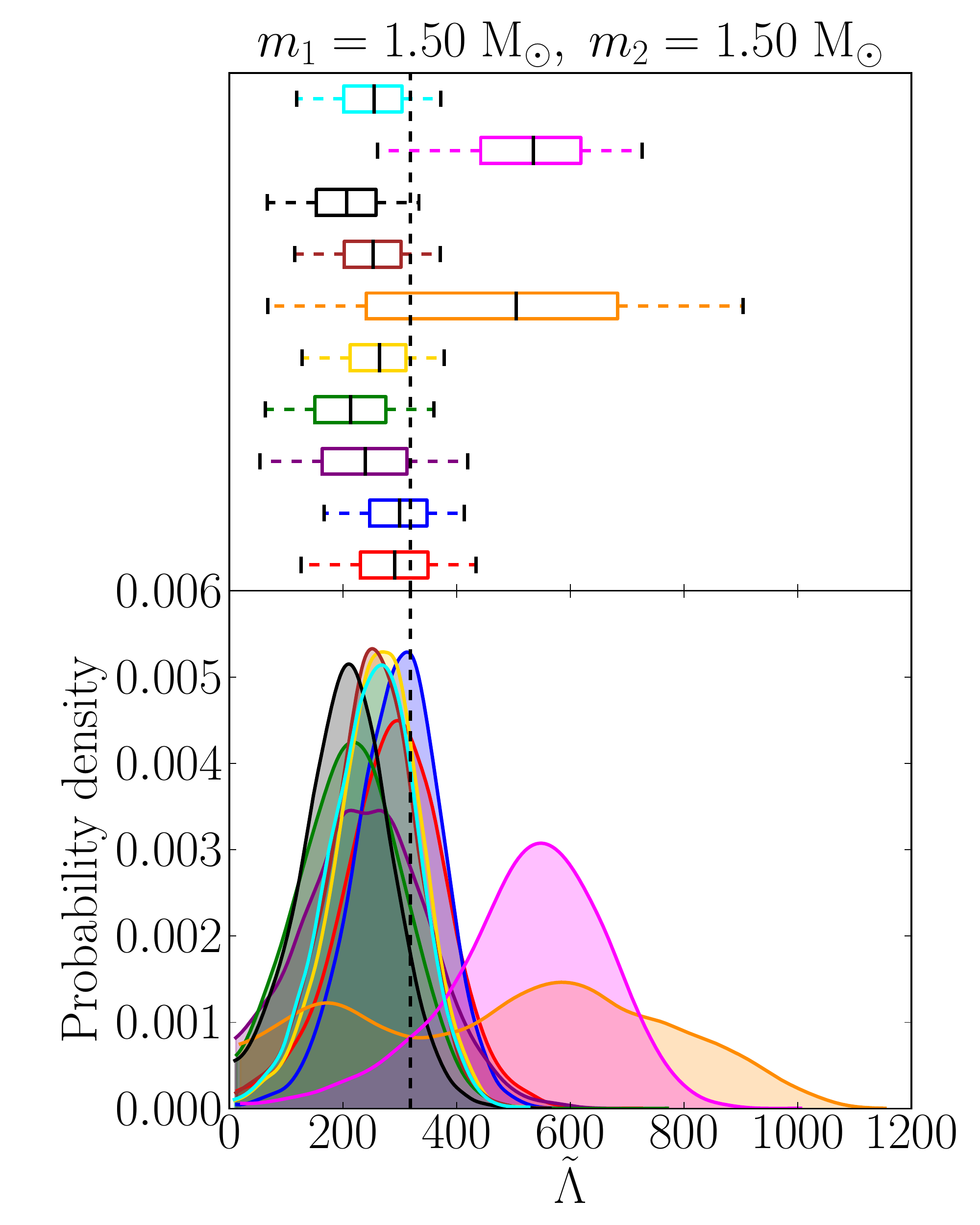}
	\end{minipage}
	~
	\begin{minipage}[b]{0.31\textwidth}
	\centering
	\includegraphics[width=\textwidth]{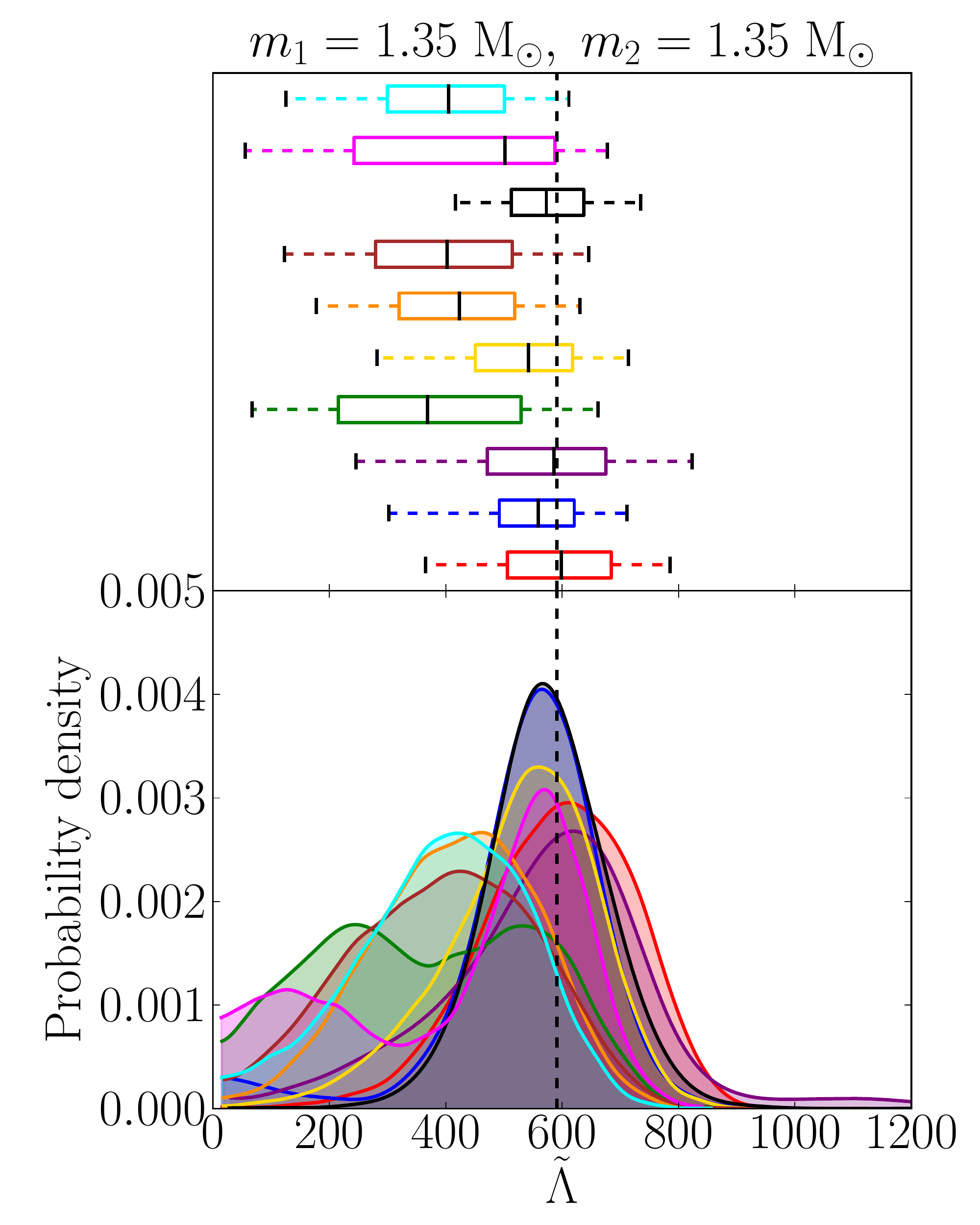}
	\end{minipage}
	~
	\begin{minipage}[b]{0.31\textwidth}
	\centering
	\includegraphics[width=\textwidth]{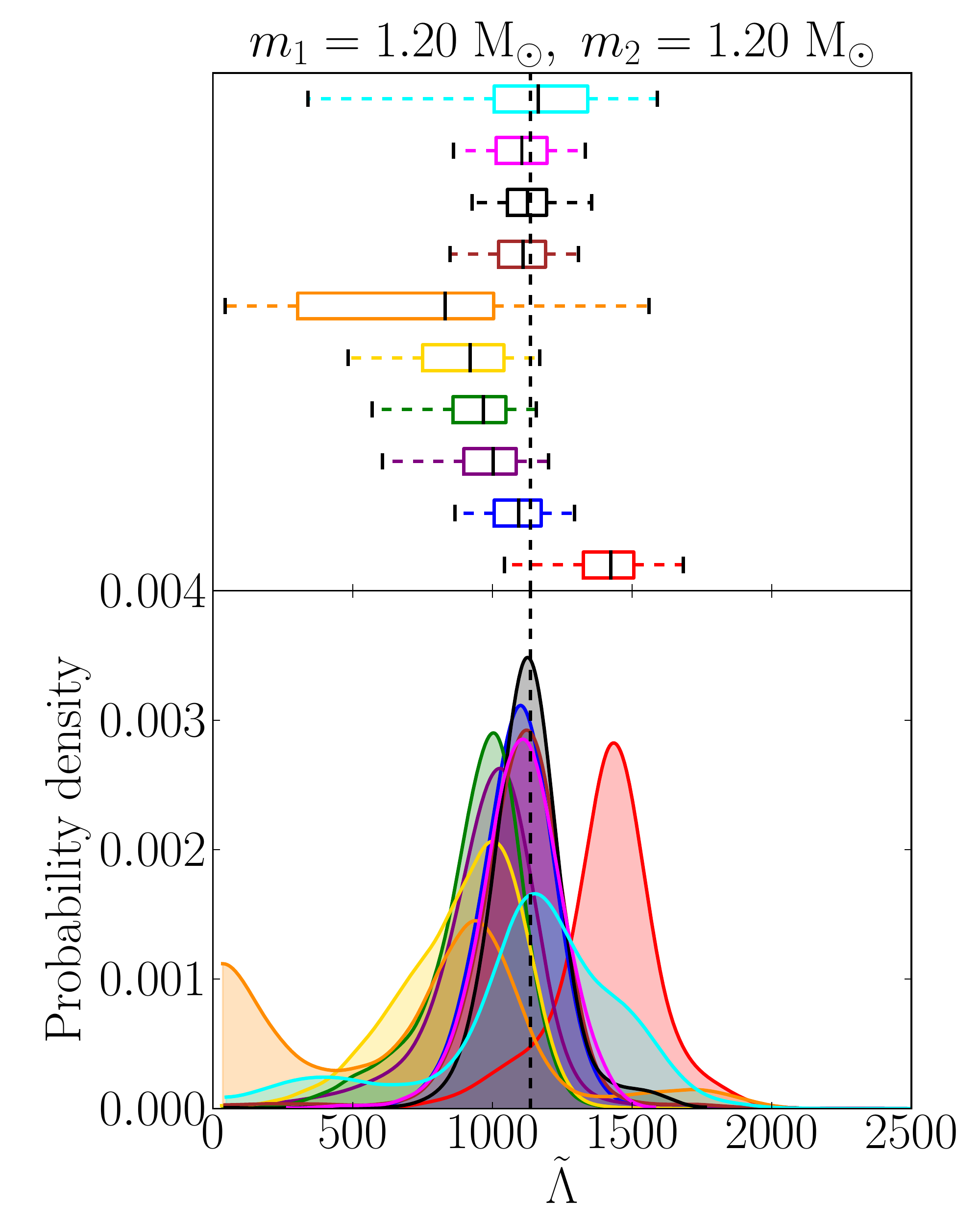}
	\end{minipage}
\caption{\label{Stat_LamT_Mc}{\footnotesize Marginalized 1D posterior probability density functions for $\tilde\Lambda$ of three BNS systems (labelled by the masses in the title) each with $\rho_{\rm net}=32.4$ (bottom).  The injected $\tilde\Lambda$ values are consistent with the MPA1 EOS model \cite{4PieceFit} and are marked with straight, dashed lines.  These plots are PDFs smoothed with a Gaussian kernel density estimator.  To generate a single plot, we inject the same BNS signal into ten different noise realizations.  The deviation of each peak away from the injected value is due to the statistical error from the presence of random detector noise.  Each PDF has an associated box-and-whisker representation (top), where the edges of each box mark the first and third quartile, the band inside each box is the median, and the end of the whiskers span the 90\% confidence interval.}}
\end{figure*}

Statistical error is due to random fluctuations in detector noise.  In Sec.~\ref{measure_tides}, all signals were injected into zero-noise, which gives the posterior averaged over noise realizations \cite{0noise}.  However, to get an understanding of how much a particular instance of noise can affect parameter recovery, we inject the same signal into ten different synthetic noise realizations (Fig.~\ref{Stat_LamT_Mc}).  Here, both the injected waveform model and the recovery waveform model is TaylorF2, and each injection has $\rho_{\rm net}=32.4$.

We find that the measurability of $\tilde{\Lambda}$ can vary dramatically from one instance of noise to the next.  A few out of the ten PDFs plotted in Fig.~\ref{Stat_LamT_Mc} have significantly broadened peaks, and some even inherit strange multimodal behavior.  Therefore, even though the true parameter value still lies within the 90\% confidence interval 90\% of the time (as expected \cite{Vallisneri}), statistical error occasionally acts to significantly reduce the measurability of $\tilde{\Lambda}$.  Unfortunately some BNS detections may provide uninformative tidal deformability estimates due to random detector noise.  Multiple detections might need to be combined to overcome the effects of noise, which was successfully shown in \cite{DelPozzo}.


\section{Conclusion/Discussion}
\label{Sec_Conclusion}

In Sec.~\ref{measure_tides}, we have shown with full Bayesian simulations that tidal deformability in BNS systems is measurable with the advanced LIGO/Virgo network (see Table~\ref{Tab_measurability}).  This is in general agreement with FM studies \cite{Damour_EOB,Read_ME,Favata} and compliments the Bayesian results shown in \cite{DelPozzo}.  For a canonical 1.35 M$_\odot$:1.35 M$_\odot$ BNS system with the moderate EOS MPA1 recovered using the advanced LIGO/Virgo network, we find that the 1$\sigma$ measurement uncertainty of $\tilde\Lambda$ (or the radius-like $\tilde\Lambda^{1/5}$) will likely be $\sim$40\% ($\sim$8\%) for a source with $\rho_{\rm net}=20$ and $\sim$20\% ($\sim$4\%) for a source with $\rho_{\rm net}=30$.

We showed in Sec.~\ref{Sec_NS_EOS} how simultaneous measurements of $\tilde\Lambda$ and chirp mass can be used to constrain the NS EOS.  Other studies in constraining the NS EOS with future GW observations include work by Del Pozzo {\it et al.}\ \cite{DelPozzo}, in which Bayesian simulations are used to incorporate information from tens of detections to discriminate between stiff, moderate, and soft EOSs.  While Del Pozzo {\it et al.}\ showed that tens of BNS sources can constrain $\lambda$ for a 1.4 M$_\odot$ NS, which can then be used to constrain the NS EOS, it might even be possible to constrain the full form of the NS EOS over all masses.

In the work presented here, we have examined the ability of GW detectors to measure the tidal parameters $\tilde\Lambda$ and $\delta\tilde\Lambda$.  The main quantity of interest, however, is the universal EOS that is common to all NSs.  One method to measure the EOS is to construct a parameterized EOS (e.g. \cite{4PieceFit,SteinerFit,LindblomFit}), then replace the tidal parameters in the waveform with EOS parameters.  This method allows one to use physical and astrophysical information to place tighter constraints on the priors for the EOS parameters in contrast to the less physically motivated priors on $\tilde\Lambda$ and $\delta\tilde\Lambda$, and this work is in preparation \cite{Wade}.  Additional work is also in progress to combine information from several BNS sources to more tightly constrain EOS parameters \cite{LackeyWade}.

Both systematic error and individual noise realizations have been shown to significantly affect the measurement of tidal deformability.  Individual instances of detector noise can severely broaden the peaks of the marginalized $\tilde{\Lambda}$ posteriors, but can be overcome by combining information from multiple sources, which averages out the effects of noise.  This would require many ($\sim$20) BNS detections \cite{DelPozzo}, instead of just a few loud signals.  Both optimistic and realistic estimates for the BNS detection rate predict that it will take less than a year after reaching design sensitivity ($\sim$2019) to constrain the NS EOS with GW signals.  However, according to pessimistic estimates, this may take considerably longer \cite{LIGO_Rates}.  Systematic error, which can significantly bias the recovered parameters, is overcome by improving current waveforms.  Higher order point-particle terms would be required in order to trust PN waveform families at frequencies sufficiently high to recover tidal deformability.  However hybrid waveforms, which are PN waveforms at low frequencies stitched to NR waveforms at high frequency, or phenomenological waveforms, which are waveforms fitted to NR, will likely be required to reliably capture high frequency effects, such as tidal deformability \cite{Read_NR,Read_ME,Lackey_NSBH,Pannarale_NSBH_WF}.  We hope that these results motivate the importance of prioritizing waveform development that incorporates NS matter effects.

\section{Acknowledgments}

L.W. would like to thank Madeline Wade, Will Farr, Chris Pankow, Justin Ellis, and Richard O'Shaughnessy for helpful discussions.  This work was partially funded by the NSF through grant numbers PHY-0970074 and PHYS-1307429, and through CAREER award number PHY-0955929.  This work required the extensive use of the Nemo computer cluster supported by the NSF under grant number PHY-0923409.

\appendix
\section{Tidally corrected PN waveform derivations}
\label{PN_derivations}

We now adopt units where $G=c=1$.  The equations that describe the CBC orbital phase evolution are the following:
\begin{eqnarray}
\label{ode_phi}
\frac{d\phi}{dt}&=&\frac{v^3}{M}\\
\label{ode_v}
\frac{dv}{dt}&=&\frac{dv}{dE}\frac{dE}{dt}=\frac{-{L}}{E^\prime},
\end{eqnarray}
where $\phi$ is the binary's orbital phase, $t$ is time, the prime represents a derivative with respect to $v$, and the requirement for energy balance is $dE/dt=-{L}$.  Integrating Eqs.~\eqref{ode_phi} and~\eqref{ode_v} give the alternate form:
\begin{eqnarray}
\label{int_t}
t(v)&=&t_{\rm ref}+\int_v^{v_{\rm ref}}\frac{E^\prime(u)}{{L}(u)}du\\
\label{int_phi}
\phi(v)&=&\phi_{\rm ref}+\int_v^{v_{\rm ref}}\frac{u^3}{M}\frac{E^\prime(u)}{{L}(u)}du,
\end{eqnarray}
where $t_{\rm ref}=t(v_{\rm ref})$, $\phi_{\rm ref}=\phi(v_{\rm ref})$, and $v_{\rm ref}$ is an arbitrary reference velocity, following \cite{BIOPS}.  Solutions for $\phi(t)$ and $v(t)$ fully determine a non-spinning CBC waveform with polarizations that go like
\begin{eqnarray}
h_+(t)&\propto&v^2 \cos 2\phi \\
h_\times(t)&\propto&v^2 \sin 2\phi.
\end{eqnarray}

Because there are several ways to solve for the orbital phase starting with the same energy and luminosity inputs, there are several different PN waveform families.  These PN families are equivalent up to unknown truncation terms at the next PN order.  We briefly outline each waveform family below and point out how tidal corrections are incorporated in their derivation.  See \cite{BIOPS} for the point-particle terms for each waveform family and details regarding initial conditions.

\subsection{TaylorT1}

The TaylorT1 approximant is achieved by numerically solving Eqs.~\eqref{ode_phi} and~\eqref{ode_v} for $\phi(t)$ and $v(t)$.  Tidal corrections enter through the energy derivative $E^\prime$ and the luminosity ${L}$:
\begin{eqnarray}
E(v)&=&E_{\rm pp}+\delta{E_{\rm tidal}}\\
E^\prime(v)&=&E_{\rm pp}^\prime+\delta{E_{\rm tidal}^\prime}\\
{L}(v)&=&{L}_{\rm pp}+\delta{L}_{\rm tidal},
\end{eqnarray}
where $\delta{E_{\rm tidal}}$ and $\delta{L}_{\rm tidal}$ come from Eqs.~\eqref{CBC_dE} and~\eqref{CBC_dEdot} respectively.

\subsection{TaylorT2}

The TaylorT2 approximant is achieved by solving Eqs.~\eqref{int_t} and~\eqref{int_phi}.  First, the ratio $E^\prime/{L}$ is expanded about $v=0$ to consistent PN order, then the result is analytically integrated to find $t(v)$ and $\phi(v)$.  Tidal corrections enter through the energy derivative $E^\prime$ and the luminosity ${L}$ and appear at 5PN and 6PN order in $t(v)$ and $\phi(v)$:
\begin{widetext}
\begin{eqnarray}
\delta\phi_{\rm tidal}(v)&=&-\frac{1}{32\eta x^{5/2}}\left[   \left(\frac{72}{\chi_1}-66\right)\frac{\lambda_1}{M^5}x^{5} +\left(  \frac{15895}{56 \chi _1}-\frac{4595}{56}-\frac{5715}{28}\chi _1+\frac{325}{14}\chi _1^2    \right)\frac{\lambda_1}{M^5}x^{6} + (1\longleftrightarrow 2)\right]\\
\delta t_{\rm tidal}(v)&=&-\frac{5M}{256\eta x^4}\left[   \left(\frac{288}{\chi_1}-264\right)\frac{\lambda_1}{M^5}x^{5} + \left(  \frac{3179}{4 \chi _1}-\frac{919}{4}-\frac{1143}{2}\chi _1+65 \chi _1^2    \right)\frac{\lambda_1}{M^5}x^{6} + (1\longleftrightarrow 2)\right].
\end{eqnarray}
\end{widetext}
Here, $x=v^2=(\pi M f_{\rm gw})^{2/3}$ is the PN expansion parameter.  The tidal corrections add  linearly to the point-particle terms:
\begin{eqnarray}
\phi(v)&=&\phi_{\rm pp}(v)+\delta{\phi_{\rm tidal}}(v)\\
t(v)&=&t_{\rm pp}(v)+\delta{t_{\rm tidal}}(v).
\end{eqnarray}
These parametric equations are then solved numerically to obtain $\phi(t)$ and $v(t)$.

\subsection{TaylorT3}

The TaylorT3 approximant starts by following the TaylorT2 approach.  After $t(v)$ and $\phi(v)$ are found, the following reparameterization is used:
\begin{equation}
\theta(t)=\left[\frac{t_{\rm ref}-t(v)}{5M}\eta\right]^{-1/8}.
\end{equation}
Next, $v(\theta)$ is found to consistent PN order via reversion of the power series.  The characteristic velocity $v(\theta)$ can then be used to find the 5PN and 6PN tidal corrections to the phase $\phi(\theta)=\phi(v(\theta))$ as well as the 5PN and 6PN tidal corrections to the GW frequency $f_{\rm gw}=v^3/(\pi M)$:
\begin{widetext}
\begin{eqnarray}
\delta\phi_{\rm tidal}(\theta)&=&-\frac{1}{\eta \theta^5}\left[   \left(\frac{9}{128\chi_1}-\frac{33}{512}\right)\frac{\lambda_1}{M^5}\theta^{10} +  \right .\\
&&\left . \left(   \frac{23325}{229376\chi _1}-\frac{12995}{1376256} -\frac{7285}{57344}\chi _1+ \frac{4885}{114688}\chi _1^2    \right)\frac{\lambda_1}{M^5}\theta^{12} + (1\longleftrightarrow 2)\right]\\
\delta f_{\rm gw,tidal}(\theta)&=&\frac{\theta^3}{8\pi M}\left[   \left(\frac{27}{256\chi_1}-\frac{99}{1024}\right)\frac{\lambda_1}{M^5}\theta^{10} +  \right .\\
&&\left .    \left(  \frac{18453}{131072\chi _1}+\frac{79}{65536}-\frac{14055}{65536}\chi _1 +\frac{171}{2048}\chi _1^2    \right)\frac{\lambda_1}{M^5}\theta^{12} + (1\longleftrightarrow 2)\right].
\end{eqnarray}
\end{widetext}
The tidal corrections add  linearly to the point-particle terms:
\begin{eqnarray}
\phi(\theta)&=&\phi_{\rm pp}(\theta)+\delta{\phi_{\rm tidal}}(\theta)\\
f_{\rm gw}(\theta)&=&f_{\rm gw,pp}(\theta)+\delta{f_{\rm gw,tidal}}(\theta).
\end{eqnarray}
These equations are essentially the equations for $\phi(t)=\phi(\theta(t))$ and $v(t)=\left[\pi M f_{\rm gw}(\theta(t))\right]^{1/3}$.

\subsection{TaylorT4}

The TaylorT4 approximant is achieved by numerically solving Eqs.~\eqref{ode_phi} and~\eqref{ode_v} for $\phi(t)$ and $v(t)$ after first expanding the ratio $E^\prime/{L}$ about $v=0$ to consistent PN order.  The 5PN and 6PN tidal corrections are:
\begin{widetext}
\begin{equation}
\delta\dot{v}_{\rm tidal}=\frac{32}{5}\frac{\eta}{M}x^{9/2}\left[   \left(\frac{72}{\chi_1}-66\right)\frac{\lambda_1}{M^5}x^{5} +   \left(    \frac{4421}{56\chi_1} - \frac{12263}{56} + \frac{1893}{4}\chi_1 - \frac{661}{2}\chi_1^2   \right)\frac{\lambda_1}{M^5}x^{6} + (1\longleftrightarrow 2)\right],
\end{equation}
\end{widetext}
where the dot represents a derivative with respect to $t$.  The tidal corrections add  linearly to the point-particle terms:
\begin{equation}
\dot{v}(v)=\dot{v}_{\rm pp}(v)+\delta{\dot{v}_{\rm tidal}}(v).
\end{equation}

\subsection{TaylorF2}

The CBC gravitational waveform can also be derived in the frequency domain using the stationary phase approximation.  The waveform takes the form
\begin{equation}
\label{psi}
\tilde{h}(f_{\rm gw})=A(f_{\rm gw})\exp\left[i\psi(f_{\rm gw})\right],
\end{equation}
where $\psi(f_{\rm gw})=2\pi f_{\rm gw} t(v) - 2\phi(v) - \pi/4$.  Substituting Eqs.~\eqref{int_t} and~\eqref{int_phi} for $t$ and $\phi$ into $\psi$ yields:
\begin{equation}
\label{int_psi}
\psi(f_{\rm gw})=2\pi f_{\rm gw} t_{\rm ref}-2\phi_{\rm ref}+2\int_v^{v_{\rm ref}}\frac{v^3-u^3}{M}\frac{E^\prime(u)}{{L}(u)}du.
\end{equation}
The tidal corrections are found by expanding the ratio $E^\prime/{L}$ about $v = 0$ to consistent PN order and integrating the expression in Eq.~\eqref{int_psi}.  By choosing to neglect amplitude corrections, the waveform becomes:
\begin{equation}
\tilde{h}(f)={\cal A}f_{\rm gw}^{-7/6}\exp\left[i\psi(f_{\rm gw})\right],
\end{equation}
where ${\cal A}\propto{\cal M}_{\rm c}^{5/6}/D$.  The chirp mass ${\cal M}_{\rm c}=\eta^{3/5}M$, and $D$ is the distance between the GW detector and the binary.  The 5PN and 6PN tidal corrections are:
\begin{widetext}
\begin{equation}
\label{F2_phase}
\delta\psi_{\rm tidal}=\frac{3}{128\eta x^{5/2}}\left[ -\left(\frac{288}{\chi _1}-264\right)\frac{\lambda_1}{M^5}x^5 - \left(\frac{15895}{28 \chi _1}-\frac{4595}{28}-\frac{5715}{14}\chi _1+\frac{325}{7}\chi _1^2\right)\frac{\lambda_1}{M^5}x^6 +  (1\longleftrightarrow 2)  \right].
\end{equation}
\end{widetext}
The tidal corrections add  linearly to the point-particle terms:
\begin{equation}
\psi(v)=\psi_{\rm pp}(v)+\delta{\psi_{\rm tidal}}(v).
\end{equation}

The TaylorF2 waveform is one of the most utilized CBC waveforms because its fully analytic frequency-domain form makes it the fastest PN waveform to generate.

\bibliography{biblio}

\end{document}